\documentclass[preprint,aps,amsmath,nofootinbib,tightenlines]{revtex4}
\usepackage{graphicx}
\usepackage{bm}
\usepackage{epsfig}
\usepackage{graphics}
\usepackage{amsfonts,amssymb}

\def\xslash#1{{\rlap{$#1$}/}}
\def\Dsl{\hbox{/\kern-.6000em D}} 

\def\dsl{\,\raise.15ex\hbox{/}\mkern-13.5mu D}
\def\bsigma{\mbox{\boldmath $\sigma$}}

\def\bsigma{\mbox{\boldmath $\sigma$}}

\def\ltap{\ \raise.3ex\hbox{$<$\kern-.75em\lower1ex\hbox{$\sim$}}\ }
\def\gtap{\ \raise.3ex\hbox{$>$\kern-.75em\lower1ex\hbox{$\sim$}}\ }
\def\OMIT#1{}

\def\lsim{\mathrel{\raise.3ex\hbox{$<$\kern-.75em\lower1ex\hbox{$\sim$}}}}
\def\gsim{\mathrel{\raise.3ex\hbox{$>$\kern-.75em\lower1ex\hbox{$\sim$}}}}

\def\msb{{\overline{\rm MS}}}

\newcommand{\nn}{\nonumber}

\newcommand{\bmk}{\mathbf k}
\newcommand{\bmp}{\mathbf p}

\newcommand{\bmq}{\mathbf q}

\newcommand{\bmI}{\mathbf I}
\newcommand{\bmJ}{\mathbf J}

\newcommand{\bmS}{\mathbf S}
\newcommand{\bmL}{\mathbf L}
\newcommand{\bmone}{\mathbf 1}

\newcommand{\bmsigma}{\mathbf \bsigma}

\newcommand{\bmx}{\mathbf x}
\newcommand{\bmy}{\mathbf y}
\newcommand{\bmn}{\mathbf n}

\newcommand{\ihat}{\,\,\!\widehat{\!\textit{\i}}}
\catcode`\@=11
\def\slash{\mathpalette\make@slash}
\def\make@slash#1#2{\setbox\z@\hbox{$#1#2$}%
  \hbox to 0pt{\hss$#1/$\hss\kern-\wd0}\box0}
\catcode`\@=12 

\begin{document}


\preprint{ \vbox{ \hbox{MPP-2006-48} 
}}

\title{\phantom{x}\vspace{0.5cm} 
Heavy Pair Production Currents 
with General  \\[3mm]
Quantum Numbers in 
Dimensionally Regularized NRQCD 
\vspace{1.0cm} }

\author{Andr\'e~H.~Hoang and Pedro~Ruiz-Femen\'\i a \vspace{0.5cm}}
\affiliation{Max-Planck-Institut f\"ur Physik\\
(Werner-Heisenberg-Institut), \\
F\"ohringer Ring 6,\\
80805 M\"unchen, Germany\vspace{1cm}
\footnote{Electronic address: ahoang@mppmu.mpg.de, ruizfeme@mppmu.mpg.de}\vspace{1cm}}


\begin{abstract}
\vspace{0.5cm}
\setlength\baselineskip{18pt}
We discuss the form and construction of general color singlet heavy
particle-antiparticle pair production currents for arbitrary quantum
numbers, and issues related to evanescent spin operators and
scheme-dependences in nonrelativistic QCD (NRQCD) in $n=3-2\epsilon$ 
dimensions. 
The anomalous dimensions of the leading interpolating
currents for heavy quark and colored scalar pairs in arbitrary 
${}^{2S+1}L_J$ angular-spin states are determined at next-to-leading
order in the nonrelativistic power counting. 
\end{abstract}
\maketitle


\newpage

%
%
%
\section{Introduction}
\label{sectionintroduction}

The use of techniques and concepts from effective field theories
(EFT's) has lead to an encouraging record of achievements for the
description of nonrelativistic fermion-antifermion systems. The
formulation of a nonrelativistic EFT for such systems has gone through
a number of important conceptual states~\cite{Reviews}. Bodwin,
Braaten and 
Lepage~\cite{Caswell,BBL} initially formulated a method for separating 
fluctuations at short distances of order the heavy particle mass $m$
from those at long distances related to the nonrelativistic momentum
and energy, $mv$ and $mv^2$ ($v$ being the relative velocity) and the
hadronic scale $\Lambda_{\rm QCD}$. Subsequent work helped to clarify
the power counting in $v$, and the relevant degrees of freedom needed
to construct an EFT. 

For very heavy quarkonium systems, characterized by the scale
hierarchy $m\gg m v\gg mv^2 > \Lambda_{\rm QCD}$ the formulation has
reached a mature state. In Ref.~\cite{Labelle} it was realized that it
is necessary to distinguish soft ($\sim mv$) and ultrasoft ($\sim
mv^2$) fluctuations and in Ref.~\cite{LM} that the original NRQCD
action~\cite{BBL} is problematic concerning the simultaneous power
counting of soft and ultrasoft terms. In particular, the Lagrangian
needs to be multipole-expanded in the presence of ultrasoft
gluons~\cite{Labelle,GR}. Technically this can be formulated by
introducing soft as well as ultrasoft gluons in the theory~\cite{LS}.

In Refs.~\cite{PS,PS2,PS3}, based on the hierarchy  $m\gg m v\gg
mv^2$, it was suggested to employ a series of EFT's, one for the
scales $m > \mu \gsim mv$ and one for $mv > \mu \gsim mv^2$, called pNRQCD
("potential" NRQCD). The pNRQCD is derived in a two-step matching
procedure where the intermediate soft matching scale and the pNRQCD
renormalization scale are independent at the field theoretic level. 
This is appropriate e.g. for static quarks where the momentum is fixed
by the quark separation $r\sim 1/mv$ and not correlated with the
dynamical energy fluctuations $E\sim mv^2$. The potential interactions
are generated from integrating out soft fluctuations at the soft
matching scale.  

In Ref.~\cite{LMR} is was pointed out that for dynamical heavy
quarkonium systems the dispersive correlation between ultrasoft energy
and soft momentum scales, $E\sim \bmp^2/m$, can be implemented
systematically at the field theoretic level by matching to the proper
EFT directly at the hard scale $\mu\sim m$ (one-step matching). The EFT,
called vNRQCD 
("velocity" NRQCD) has a strict power counting in $v$.
It contains soft and ultrasoft degrees of freedom as well as soft and
ultrasoft renormalization scales, $\mu_S$ and $\mu_U$. Through the
different velocity counting of soft and ultrasoft fields they are
correlated as $\mu_U\propto \mu_S^2/m$, where usually the constant of
proportionality is set to $1$, i.e. $\mu_U=\mu_S^2/m=m\nu^2$. The
renormalization group running of the theory is then conveniently
expressed in terms of the dimensionless parameter $\nu$. The matching
is carried out at the hard scale $\nu=1$ and the theory is evolved to
$\nu\sim v\sim \alpha_s$ of order of the relative velocity of the
two-body system for computations of matrix elements. The running
properly sums logarithms of the momentum and the energy scale at the
same time~\cite{LMR,amis4,mss1,HoangStewartultra} 
and is referred to as the velocity renormalization group
(VRG)~\cite{LMR}. Within dimensional regularization the powers of
$\mu_S^\epsilon$ and $\mu_U^\epsilon$ multiplying the 
operators of the Lagrangian are uniquely determined by the $v$
counting and the dimension of the operators in $d=4-2\epsilon$
dimensions. 

In Ref.~\cite{Hoang3loop}, where a three-loop renormalization of
quark-antiquark vertex diagrams was carried out for fermion-antifermion
S-wave states, is was demonstrated that the one-step matching principle, upon
which vNRQCD is based, is consistent under renormalization at the subleading
order level, when subdivergences need to be subtracted and when UV
divergences from heavy quark-antiquark loops and from soft or
ultrasoft gluons appear simultaneously in single diagrams. Up to now there is 
no analogous demonstration for the two-step matching principle. 

With the VRG, the running of potentials relevant for the
next-to-next-to-leading logarithmic (NNLL) description of the
nonrelativistic dynamics of pairs of heavy quarks and colored scalars
has been determined in Refs.~\cite{amis,amis2,amis3,HoangStewartultra}
and \cite{HoangRuizSvNRQCD}. The NLL running of leading (in the $v$
expansion) S-wave currents for heavy quark-antiquark production and of
the leading S- and P-wave currents for pairs of colored scalars were
determined in Refs.~\cite{amis3,HoangStewartultra} and
\cite{HoangRuizSvNRQCD}. Corresponding work in the pNRQCD formalism
for heavy quarks was carried out in
Refs.~\cite{PSstat,Pineda1,Pineda2,Pineda3}. In these computations,
diagrams which simultaneously contain both UV divergences from heavy
quark-antiquark loops and from soft or ultrasoft gluon loops do not arise. The
final results based on the VRG and on pNRQCD agree, after the
scale correlations from the one-step matching procedure of vNRQCD are
imposed in pNRQCD. Concerning the NNLL running of 
the production currents, only the contributions from three-loop vertex
diagrams for the leading currents describing quark-antiquark pair
production in an S-wave configuration~\cite{Hoang3loop} are 
known at this time in vNRQCD. For the NNLL contributions that arise from the 
subleading order evolution of the coefficients that appears in the NLL
anomalous dimension of the current at present only the results for the
Coulomb potential~\cite{PSstat,HoangStewartultra}, the spin-dependent
$1/m^2$ potentials~\cite{Peninspin1,Peninspin2} and from
ultrasoft quark loops~\cite{Stahlhofen1} have been determined.  

The main phenomenological application of the nonrelativistic EFT for
the situation $m\gg mv\gg mv^2>\Lambda_{\rm QCD}$ is top quark pair
production close to threshold at a future Linear Collider~\cite{LC}. 
Here the summation of logarithms of $v$ is crucial to control the
large normalization uncertainties~\cite{hmst,hmst1} that arise in 
fixed-order nonrelativistic computations that only account for the
summation of Coulomb potential insertions~\cite{HT1,Hoangsynopsis}. One
can expect that the summation of logarithms of the velocity is also
important for the description of squark pair production at
threshold~\cite{HoangRuizSvNRQCD}. 
Recently, it was also found that it is crucial to apply
the EFT for predictions of the $e^+e^-\to t\bar t H$ cross section at 
$\sqrt{s}=500$~GeV~\cite{Farrell1,Farrell2} since here the $t\bar t$
final state interactions are dominated by nonrelativistic dynamics.

In this work we address the form and construction of non-relativistic
interpolating currents for production and annihilation of a color
singlet heavy particle-antiparticle pair with general quantum
numbers ${}^{2S+1}L_J$ in $n=d-1=3-2\epsilon$ dimensions for quarks
and colored scalars within NRQCD. In general, the interpolating
currents for a specific ${}^{2S+1}L_J$ state are associated to
irreducible tensor representations of the SO(n) rotation group that
can be built from  
the particle-antiparticle relative momentum and the bilinear
covariants of the particle-antiparticle field operators. 
While in three dimensions the basis of interpolating currents for the 
different ${}^{2S+1}L_J$ states is, by construction, unique, in
general $n\neq 3$ dimensions for fermions there exist evanescent
operator structures that make the choice of basis for the
interpolating currents ambiguous. This is because the structure of
irreducible SO(n) representations for $n\neq 3$ is inherently
different and in general richer than in three dimensions. Technically
the ambiguity is related to the number of sigma 
matrices that generate the SO(n) rotations for the spinor wave
functions. For explicit computations a specific choice of basis has
to be made, which is associated to a specific choice of a
renormalization scheme. A very similar problem has been treated
systematically already some time ago in the framework of relativistic
QCD and the effective weak
Hamiltonian~\cite{DuganGrin,HerrlichNierste,ChetyrkinMisiak}, and much
of the statements that apply for the relativistic case can be directly
transferred to the nonrelativistic theory. However, due to the 
nonrelativistic power counting some even more specific statements,
can be made. In particular, 
the NLL order matching conditions and anomalous dimensions of the
${}^{2S+1}L_J$ interpolating currents that are leading in the
nonrelativistic expansion are scheme independent.

Using the explicit form of the interpolating currents obtained in this
work we also determine the NLL anomalous dimensions of the leading (in
the $v$ expansion) currents describing color singlet heavy 
quark-antiquark and squark-antisquark pair production for arbitrary
spin and angular momentum ${}^{2S+1}L_J$ configurations. The results
for low angular 
momentum states (S,P,D) are e.g. relevant for angular distributions at
the threshold or for production and decay rates of squark-antisquark
resonances in certain supersymmetric scenarios where squarks can
have a very long lifetime. The results also shed some light on the
importance of summing logarithms of 
$v$ for the production and annihilation rates of high angular momentum
states. In this respect our results help to complete analogous higher
order considerations in previous literature for the energy
levels~\cite{hms1} and the
wave-functions~\cite{Pineda2,Melnikov:1998ug}.

For the presentations in this work we employ the
notations from vNRQCD based on a label formalism for soft fields and
the heavy quarks (or scalars)~\cite{LMR}. We note, however, that  
most of the results obtained in this work are applicable to NRQCD in
general. In particular, the NLL order anomalous dimensions obtained
here are also valid in pNRQCD once the vNRQCD scale correlations from
the one-step matching are imposed. 

The outline of the paper is as follows: In Sec.~\ref{sectionsinglet}
we discuss the form of the spherical harmonics in $n$ dimensions and
present the form of currents describing states with arbitrary angular
momentum $L$ and total spin zero. We also present the $n$-dimensional
form of the Legendre polynomials and demonstrate in an example why the
$n$-dimensional form of the spherical harmonics is needed in
dimensional regularization. 
In Sec.~\ref{sectiontriplet} we discuss the properties of
$\sigma$-matrices for $n\neq 3$ and the importance of evanescent
operator structures that can be built from the $\sigma$-matrices.
We present a simple basis of interpolating currents describing
fermion-antifermion pairs in a spin triplet state and having arbitrary
$L$ and total angular momentum $J$. 
In Sec.~\ref{sectionNLLsinglet} we determine the anomalous dimension
of the current for spin singlets, and in 
Sec.~\ref{sectionNLLtriplet} those of  the current for spin triplets.
Section~\ref{sectionNLLtriplet} also contains a discussion on the
scheme-dependence of the spin-dependent potentials.
In Sec.~\ref{sectiondiscussion} we determine and solve the resulting
anomalous dimensions for heavy quarks and colored scalars. The results
are analysed numerically for a few cases.
In Sec.~\ref{sectioncomments} we comment on the scheme-dependence of
results that can be found in previous literature. 
The conclusions are given in Sec.~\ref{sectionconclusion}.
We have added a few appendices: Appendices~\ref{appendix2} and
\ref{appendix3} contain the derivation of the tree level NRQCD
matching conditions for the well known processes $q\bar q\to 2\gamma$
and $3\gamma$ accounting properly for the existence of evenescent 
operator structures. In App.~\ref{appendix4} some
more details are given on the determination of the spin triplet
currents shown in Sec.~\ref{sectionNLLtriplet}. We also derive an
alternative set of currents that is equivalent for $n=3$ but 
inequivalent for $n\neq3$. Finally in App.~\ref{appendix5} we give 
results for the UV-divergences for the elementary integrals
that are needed for the determination of the anomalous dimensions
of the currents.

\section{Spin Singlet Currents} 
\label{sectionsinglet}

In this section we discuss the form of the interpolating currents
describing the production of a particle-antiparticle pair with total
spin zero for arbitrary relative angular momentum $L$
(${}^{2S+1}L_J={}^1L_L$). They are relevant for pairs of scalars or for
quark-antiquark pairs in a spin singlet state. The currents naturally
have their simplest form in the c.m.\,frame where relevant angular
dependence can only arise from the c.m.\,spatial momenta of the
pair. Thus the generic structure of the production currents is
\begin{eqnarray}
  \psi_{\bmp}^\dagger(x)\,\Gamma(\bmp)\,\tilde \chi_{-\bmp}^*(x)
\,,
\end{eqnarray}
where $\Gamma(\bmp)$ represents an arbitrary tensor depending on the
c.m.\,momentum label $\bmp$. For the case of scalars  $\psi_{\bmp}$ and
$\tilde\chi_{\bmp}$ are scalar fields with
$\psi_{\bmp}^\dagger=\psi_{\bmp}^*$, while for fermions $\psi_{\bmp}$ and
$\chi_{\bmp}$ are Pauli spinor fields with $\tilde\chi_{-\bmp}^*\equiv
(i\sigma_2)\chi_{-\bmp}^*$. (Note that we adopt the 
standard nonrelativistic spinor convention of antiparticles with {\it
positive} energies. In this convention fermions and antifermions have
the same spin operators.) The corresponding
annihilation current can be obtained by hermitian conjugation. 
The interpolating currents associated to a definite angular momentum
state $L$ are related to irreducible representations of the tensor
$\Gamma$ with respect to the rotation group SO(n). Since the transformation
properties of the fields are not relevant in this respect for the spin
zero case, it is sufficient to identify the irreducible tensors that
can be built from the spatial momentum vector $p^i$, $i=1,\ldots,n$,
where $n=d-1$. The irreducible tensors associated to the
angular momentum $L$ are up to normalization just the spherical
harmonics of degree $L$. In the following we give a general discussion
in $n$ dimensions. All results reduce to the well known results for $n=3$.

\subsection*{Spherical Harmonics}

The spherical harmonics of degree $L$ are the
polynomials $u(\bmx)$ in $\mathbb{R}^n$ which are homogeneous of degree $L$
(i.e. $u(r \bmx)=r^L u(\bmx)$), harmonic (i.e. they satisfy the Laplace equation 
$\Delta_{\mathbb{R}^n}u(\bmx)\equiv\nabla^2 u(\bmx)=0$) and restricted to the unit
sphere  $S^{n-1}$. The 
spherical harmonics ${Y}_{LM}(n,\bar\bmx)$ of degree $L$, with
$M=1,\dots,n_L$ and $\bar\bmx=\bmx/|\bmx|$, form an orthogonal basis of a
$n_L$-dimensional vector space with
\begin{equation}
n_L\,=\,
\left(\begin{array}{c}n+L-1\\ L\end{array}\right)
-\left(\begin{array}{c}n+L-3\\ L-2\end{array}\right)
\, = \,
(2L+n-2)\,\frac{\Gamma(n+L-2)}{\Gamma(n-1)\Gamma(L+1)}\,.
\label{nL}
\end{equation}
Any differentiable function on the unit sphere $S^{n-1}$ in
$\mathbb{R}^n$ can be expanded in terms of a convergent series of
spherical harmonics.  
A representation in terms of cartesian coordinates~\footnote{
Although it is possible to write down results with spherical
coordinates for integer $n$ (see e.g. Ref.~\cite{Normand:1981dz}), it
is more convenient in practice and for actual 
computations to use cartesian coordinates. 
} 
of the spherical
harmonics of degree $L$ is given by the totally symmetric and traceless
tensors with $L$ indices 
$T^{i_1\dots i_L}(\bmx)$, where the indices $i_1,\ldots,i_L$ are cartesian
coordinates~\cite{grouptheory}. The restriction to the unit sphere is not
essential regarding the transformation properties under SO(n) rotations and
can be dropped for our purposes. An explicit expression for 
$T^{i_1\dots i_L}(x)$ reads: 
\begin{eqnarray}
T^{i_1\dots i_L}(\bmx) 
 &=& x^{i_1}\ldots x^{i_L}
   -\frac{\bmx^2 \, \Phi_1^{i_1\dots i_L}(\bmx)}{(2L+n-4)}
   +\frac{(\bmx^2)^2 \, \Phi_2^{i_1\dots i_L}(\bmx)}{(2L+n-4)(2L+n-6)}
   - \dots
\nn\\ & & 
\dots + (-1)^{[L/2]}
   \frac{(\bmx^2)^{[L/2]} \, \Phi_{[L/2]}^{i_1\dots i_L}(\bmx)}
         {(2L+n-4)\dots(2L+n-2-2[L/2])}
\nn\\[3mm] 
 &=&  
 \sum_{k=0}^{[L/2]}\, C_k^L \, \bmx^{2k}\, \Phi_k^{i_1\dots i_L}(x) \;,
\nn\\[3mm]  
C_k^L &=& (-1)^k \, 2^{-k} \frac{\Gamma(\frac{n}{2}+L-k-1)}{\Gamma(\frac{n}{2}+L-1)} \;,
\nn\\[3mm]  
\Phi_k^{i_1\dots i_L}(\bmx) & = & 
  \sum_{\mathrm{inequ. \,\,permut.}\atop(p_1\dots p_L)\,\mathrm{of}\, (1\dots L) }
  \underbrace{ \,\delta^{i_{p_1}i_{p_2}}\dots
 \delta^{i_{p_{(2k-1)}}i_{p_{2k}}} }_{k \;\delta'\mathrm{s}} 
 \, x^{i_{p_{(2k+1)}}}\dots x^{i_{p_L}} 
\nn\\ & = &
 \delta^{i_1i_2}\dots\delta^{i_{2k-1}i_{2k}}\,x^{i_{2k+1}}\dots x^{i_L}
 \, + \, \mbox{(inequ. permut.'s)}
\,,
\label{Tdef}
\end{eqnarray}
where the symbol $[\ldots]$ denotes the Gauss bracket. 
The sum in the $\Phi_k$ tensors extends over all sets of $k$ pairs of
indices that can be constructed  
from $L$ indices ($L\ge 2k$); the number of terms in the sum is thus 
${L\choose 2k}(2k-1)!!$. Note that we call two
permutations equivalent if they lead to the same term in the sum.
Although we believe that the expression for $T^{i_1\dots i_L}(x)$ has been shown somewhere in the
literature before, we were not able to locate a suitable reference.  
A few simple examples are
\begin{eqnarray} &&
T^i(\bmx) \, = \, x^i\,,\qquad
T^{ij}(\bmx) \, = \, x^ix^j-\frac{\bmx^2}{n}\delta^{ij}\,,
\nn\\ & &
T^{ijk}(\bmx) \, = \,  x^ix^jx^k-\frac{\bmx^2}{n+2}
   \left(x^i\delta^{jk}+x^j\delta^{ik}+x^k\delta^{ij}\right)\,.
\end{eqnarray}
From the Laplacian 
\begin{eqnarray}
\Delta_{\mathbb{R}^n} & \equiv & 
\frac{\partial^2}{\partial  x^{i\,2}} \, = \,
\left(\frac{\partial^2}{\partial r^2} +
  \frac{n-1}{r}\frac{\partial}{\partial r}\right) -
\frac{1}{r^2}\,\bmL^2
\,,
\nonumber\\[2mm]
-\bmL^2 & = & 
\frac{\partial}{\partial x^j}x^k\frac{\partial}{\partial x^j} x^k-
\frac{\partial}{\partial x^j} x^k \frac{\partial}{\partial x^k} x^j
\, = \,
\bmx^2\nabla^2 - 
x^k x^j \frac{\partial}{\partial x^k}\frac{\partial}{\partial x^j} -
(n-1) x^k \frac{\partial}{\partial x^k}
\,,
\end{eqnarray}
one can find an explicit form for the squared angular momentum
operator and, using the homogeneity relation 
$\Delta_{\mathbb{R}^n}T^{i_1\dots i_L}=0$, the eigenvalue equation
\begin{eqnarray}
\bmL^2\,\, T^{i_1\dots i_L}(\bmx) & = & L(L+n-2)\,\, T^{i_1\dots i_L}(\bmx)
\,.
\end{eqnarray}
We define the interpolating currents that describe production of a
spin singlet and angular momentum $L$ state (${}^{2S+1}L_J={}^1L_L$)
as
\begin{eqnarray}
(j_L^{S=0})^{i_1\ldots i_L} & \equiv & 
\psi_{\bmp}^\dagger(x)\, T^{i_1\dots i_L}(\bmp)\,\tilde
\chi_{-\bmp}^*(x)
\,.
\label{singletcurrent1}
\end{eqnarray}
The current in Eq.~(\ref{singletcurrent1}) is the dominant ${}^1L_L$
current in the $v$-expansion. Higher order currents are obtained by
including additional powers of the scalar term $\bmp^2$.  

As an example, let us consider the currents with angular momenta $S,\,P$ and $D$.
They are relevant in the electromagnetic production of
heavy charged scalars from $e^+e^-$ and $\gamma\gamma$ collisions. 
The $e^+e^-$ annihilation at lowest 
order in the expansion of the electromagnetic coupling produces $P$-wave states:
\begin{eqnarray}
i{\cal M}_{e^+e^-} =
-i\,\frac{8\pi\alpha}{s}\,Q_q \,\bar{v}(k^\prime)\,\gamma^i\, u(k)
\,T^i(\bmp)
\,,
\label{pwave}
\end{eqnarray}
with $k,\,k^\prime$ the momenta of the incoming $e^+e^-$ in the c.m.\,frame 
and $Q_q$ denoting the electromagnetic
charge of the scalars. Note that potential color indices are always implied.
Pairs of heavy scalars in $S$- and $D$-waves are first
produced in  $\gamma\gamma$ collisions. The first terms of the amplitude in the
expansion in the c.m.\,three-momentum of the outgoing particles read:
\begin{eqnarray}
i{\cal M}_{\gamma\gamma} &=&
32 i\,\pi\alpha\,Q_q^2 \,\epsilon_1^i\epsilon_2^j\,
\left[\,-\frac{1}{4}\,\delta^{ij}+\frac{p^i p^j}{2m^2} 
+ \dots \,\right]
\nn\\[3mm]
&=& -8i\,\pi\alpha\,Q_q^2 \,\epsilon_1^i\epsilon_2^j\left(
1-\frac{2\,\bmp^2}{n \,m^2}
\right) +
\frac{16i\,\pi\alpha}{m^2}\,Q_q^2 \,
\epsilon_1^i\epsilon_2^j \, T^{ij}(\bmp)
+ \dots\,.
\label{s-d-waves}
\end{eqnarray}
The first term in the last equality is the $S$-wave
contribution ($T=1$) whereas the second is the $D$-wave contribution described
by the spherical harmonic of degree 2, $T^{i_1i_2}(\bmp)$.

Some useful relations for the $T$ tensors read:
\begin{eqnarray}
T^{\,i_1\ldots i_L}(\bmx)\, x^{i_L}
 &=& \frac{L+n-3}{2L+n-4}\,\bmx^2\,T^{\,i_1\ldots i_{L-1}}(\bmx) \;,
 \nn\\[3mm] 
T^{\,i_1\ldots i_L}(\bmx)\,  x^{i_{L+1}} 
 &=&  T^{\,i_1\ldots i_{L+1}}(\bmx) 
     - \frac{2 \,\bmx^2}{(2L+n-2)(2L+n-4)}\, \sum_{j< k \atop j=1}^L \, \delta^{i_j i_k}   \,  
     T^{\,i_1\dots\ihat_j\dots\ihat_k\dots i_{L+1} } (\bmx)
 \nn\\[1mm]    
 && +\;\frac{\bmx^2}{2L+n-2} \, \sum_{j=1}^L \, \delta^{i_j i_{L+1}} \, 
     T^{\,i_1\dots\ihat_j\dots i_{L} } (\bmx) \;,
 \nn\\[3mm]   
\frac{\partial}{\partial x^\ell}\,\Phi_k^{i_1\dots i_L}(\bmx) &=&
 \sum_{j=1}^L \, \delta^{i_j\ell}\,\Phi_{k}^{i_1\dots\ihat_j\dots i_L}(\bmx)
\, = \,
 \Phi_{k+1}^{i_1\dots i_L \ell}(\bmx) - x^\ell\,\Phi_{k+1}^{i_1\dots i_L}(\bmx) \;,
 \nn\\[3mm]
x^\ell\,\Phi_{k}^{i_1\dots i_L}(\bmx) &= &
   \Phi_{k}^{i_1\dots i_L \ell}(\bmx) - 
   \sum_{j=1}^L\,\delta^{i_j\ell}\,\Phi_{k-1}^{i_1\dots\ihat_j\dots i_L}(\bmx) \;,
\nn\\[3mm]
x^{[\ell}\,\partial^{\,k]}_x\,T^{\,i_1\ldots i_L}(\bmx) &=& 
  \frac{1}{2}\sum_{j=1}^{L} \,  \delta^{i_j k}  \,
     T^{\,i_1\dots\ihat_j\dots i_{L} \ell} (\bmx) - \{ \ell \leftrightarrow   k \}\;,   
\label{Trelations}
\end{eqnarray}
where we use the notation that hatted indices are dropped, and 
$A^{[ij]}\equiv\frac{1}{2}(A^{ij}-A^{ji})$.

\subsection*{Legendre Polynomials}

The spherical harmonics of degree $L$ satisfy the addition theorem
\begin{equation}
\sum_{M=1}^{n_L} {Y}_{LM}(n,\bar \bmx) \, {Y}_{LM}^*(n,\bar
\bmy)=\frac{n_L}{\sigma_n}\,P_L(n,\bar \bmx.\bar\bmy) 
\,,
\label{addtheorem1}
\end{equation}
where $P_L(n,t)$, $t\in[-1,1]$, is the Legendre polynomial
of degree  $L$ generalized to $n$ dimensions, with $P_L(n,1)=1$ and
$\sigma_n=\frac{2\pi^{n/2} }{\Gamma(n/2)}$ is the area of the unit sphere $S^{n-1}$. The
Legendre polynomial is related to the Gegenbauer polynomial
$C_L^{\lambda}(t)$.\footnote{
  The index $\lambda$ is
  chosen such that the $C_L^{\lambda}$ are orthogonal on the
  $n$-dimensional sphere
} 
One has
$P_L(n,t)=\frac{\Gamma(n-2)\Gamma(L+1)}{\Gamma(L+n-2)}C_L^{\lambda}(t)$,
where $\lambda=\frac{n-2}{2}$.
The explicit form of the generalized Legendre polynomial
reads~\cite{Gradstheyn},  
\begin{equation}
P_L(n,t)=\frac{\Gamma(n-2)\,\Gamma(L+1)}{\Gamma(\frac{n-2}{2})\, 
\Gamma(n+L-2)}\,\sum_{k=0}^{\left[\frac{L}{2}\right]}
\,\frac{(-1)^k\,\Gamma(\frac{n-2}{2}+L-k)}{\Gamma(k+1)\,\Gamma(L-2k+1)}
\;(2t)^{L-2k} 
\;.
\label{Legendredef}
\end{equation}
From Eq.~(\ref{addtheorem1}) follows an addition theorem for the $T$ tensors
upon total contraction of their indices:
\begin{equation}
T_{L,n}(\bmx,\bmy)
\, \equiv \,
T^{\,i_1\dots i_{L}}(\bmx)\, T^{\,i_1\dots i_{L}}(\bmy) 
\, = \, 
|\bmx|^L\,|\bmy|^L\,N_L \, P_L(n,t) \;,
\qquad t=\frac{\bmx.\bmy}{|\bmx||\bmy|}\;,
\label{Tcontract}
\end{equation}
where
\begin{equation}
N_L \, \equiv \,
\frac{\Gamma(L+n-2)\,\Gamma(\frac{n-2}{2})}{2^{L}\,\Gamma(L+\frac{n}{2}-1)\,\Gamma(n-2)}\,.
\end{equation}
Another very useful contraction is:
\begin{eqnarray}
\lefteqn{
T^{\,i_1\dots i_{L-1}j}(\bmx)\,T^{\,i_1\dots
  i_{L-1}\ell}(\bmy) \,\left(y^j x^{\ell}-x^j y^\ell \right) 
\,=\,
}
\nn\\ & &
|\bmx|^{L+1}\,|\bmy|^{L+1}\,
N_L \,
\frac{L+n-2}{2L+n-2}\,\left[ P_{L+1}(n,t) -P_{L-1}(n,t)  \right]
\;.
\label{TyTx}
\end{eqnarray}

\subsection*{Consistency Consideration}

The use of the generalized currents in
Eqs.~(\ref{Tdef}) and~(\ref{singletcurrent1}) is mandatory to obtain
consistent results in dimensional regularization in accordance with
SO(n) rotational invariance. As an example let us consider the
nonrelativistic three-loop vacuum polarization diagram shown in
Fig.~\ref{fig:3loopCoul} with two insertions of the Coulomb potential.
Inserting the currents that produce and annihilate the
particle-antiparticle pair with angular momentum $L$ and fully
contracting the indices, the quantity we want to compute is, up to a
global factor 
($D^n\bmp\equiv \tilde{\mu}_s^{2\epsilon}  d^n\bmp/(2\pi)^{n},\;
\tilde{\mu}_s^{2\epsilon}=\mu_s^{2\epsilon} (4\pi)^{-\epsilon} e^{\epsilon\gamma}$):
%
%
\begin{figure}[t] 
\begin{center}
 \leavevmode
 \epsfxsize=6cm
 \leavevmode
 \epsffile{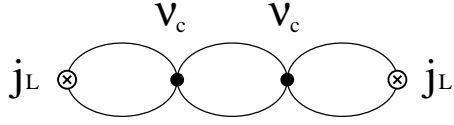}
 \vskip  0.0cm
 \caption{Three-loop diagram with two insertions of the Coulomb
 potential (black dots). 
 \label{fig:3loopCoul} }
\end{center}
\end{figure}
\begin{eqnarray}
\frac{1}{N_L}\int D^n \bmq_1\, D^n \bmq_2\, D^n \bmq_3 \,
\frac{T^{\,i_1\dots i_{L}}(\bmq_1)\, T^{\,i_1\dots i_{L}}(\bmq_3) }
{(\bmq_1^2+\delta)\,(\bmq_1-\bmq_2)^2\,(\bmq_2^2+\delta)\,
(\bmq_2-\bmq_3)^2\,(\bmq_3^2+\delta)}\,,  
\end{eqnarray}
with $\delta=-mE-i\epsilon$. We can now proceed in two different ways
to do the computation. From Eq.~(\ref{Tcontract}), the contraction of
the $T$'s at the ends generates a Legendre polynomial  
depending on the angle between the loop momenta on the sides:
\begin{eqnarray}
{\cal A}_1^{(L)} = \int D^n \bmq_1\, D^n \bmq_2\, D^n \bmq_3 \,\frac{|\bmq_1|^L |\bmq_3|^L \,
P_L(n,\bar \bmq_1 . \bar \bmq_3) }
{(\bmq_1^2+\delta)\,(\bmq_1-\bmq_2)^2\,(\bmq_2^2+\delta)\,
(\bmq_2-\bmq_3)^2\,(\bmq_3^2+\delta)}\,.
\label{A1}
\end{eqnarray}
On the other hand, we can first shift the loop momenta dependence of
the current $T^{\,i_1\dots i_{L}}(\mathbf q_1)$ by using the 
relation
\begin{eqnarray}
 &&\int d^n \bmq_1 \, T^{\,i_1\dots i_{L}}(\bmq_1)\,f( \bmq_1, \bmq_2)= A\,
  T^{\,i_1\dots i_{L}}(\mathbf q_2)\,,
\nn\\[2mm]
&&
A= \int d^n \bmq_1 \,|\bmq_1|^L |\bmq_2|^{-L} \,
P_L(n,\bar \bmq_1 . \bar \bmq_2) \,f( \bmq_1, \bmq_2) \,,
\label{shift}
\end{eqnarray}
which is a consequence of rotational invariance if 
$f( \bmq_1, \bmq_2)$ is a scalar function depending on
$\bmq_1^2,\,\bmq_2^2,\,\bmq_1\cdot\bmq_2$. 
This gives an alternative expression
with two Legendre polynomials:
\begin{eqnarray}
{\cal A}_2 ^{(L)} = 
\int D^n \bmq_1\, D^n \bmq_2\, D^n \bmq_3 \,
\frac{|\bmq_1|^L |\bmq_3|^L \,
P_L(n,\bar\bmq_1.\bar\bmq_2)\,P_L(n,\bar\bmq_2.\bar\bmq_3)}
{(\bmq_1^2+\delta)\,(\bmq_1-\bmq_2)^2\,(\bmq_2^2+\delta)\,
(\bmq_2-\bmq_3)^2\,(\bmq_3^2+\delta)}\,.  
\label{A2}
\end{eqnarray}
Had we worked in $n=3$ from the beginning, we would have written down
Eqs.~(\ref{A1}) and~(\ref{A2}) with 
the generalized Legendre polynomials replaced by their $n=3$
expressions. Let us call the corresponding 
expressions as $\tilde{{\cal A}}_1^{(L)},\;\tilde{{\cal A}}_2^{(L)}$. 
Since the integrals are 
power divergent, we can compute them using dimensional regularization in $n=3-2\epsilon$
dimensions and check whether they give the same result. 
The first non-trivial case is $L=2$, because one has $P_{0,1}(3,x)=P_{0,1}(n,x)$. 
The results for $\tilde{{\cal A}}_1^{(2)},\;\tilde{{\cal A}}_2^{(2)}$ then read
\begin{eqnarray}
\tilde{{\cal A}}_1^{(2)} &=& \frac{2^{-6\epsilon}}{3072\pi^3}\,
(\delta)^{\frac{3n}{2}-3}\mu_s^{6\epsilon}\,
\Big\{15+4\pi^2+\epsilon\left[32\pi^2-120\,\zeta(3)
 +231\right]+{\cal O}(\epsilon^2) \Big\}\,,
\nn\\[3mm]
\tilde{{\cal A}}_2^{(2)} &=& \frac{2^{-6\epsilon}}{3072\pi^3}\,
(\delta)^{\frac{3n}{2}-3}\mu_s^{6\epsilon}\,
\Big\{
18+4\pi^2+\epsilon\left[ 32\pi^2-120\,\zeta(3)
 +264-4\log 8 \right]+{\cal O}(\epsilon^2)\Big\}\,.
\nn\\
\end{eqnarray}
A difference arises even in the first term in the $\epsilon$
expansion, which reflects the fact that 
$\tilde{{\cal A}}_1^{(2)},\;\tilde{{\cal A}}_2^{(2)}$ are not the same
quantity. The reason is that the shift (\ref{shift}) performed in
$n=3$ dimensions and the following evaluation using dimensional
regularization do not commute if the integrals are divergent. 
Rotational invariance in $n$ dimensions
must be satisfied at every step if dimensional regularization is
used. This is automatically achieved by using the SO(n) tensors $T$ as
shown in Eqs.~(\ref{A1}) and~(\ref{A2}). The computation 
with generalized Legendre polynomials corresponding to ${\cal
  A}_1^{(L)}$ and ${\cal A}_2^{(L)}$ in Eqs.~(\ref{A1},\ref{A2}) can
be shown to produce the correct outcome. For $L=2$ the correct result
reads 
\begin{eqnarray}
{\cal A}_1^{(2)} = {\cal A}_2^{(2)} =\frac{2^{-6\epsilon}}{3072\pi^3}\,
(\delta)^{\frac{3n}{2}-3}\mu_s^{6\epsilon}\,
\Big\{
15+4\pi^2+\epsilon\left[ 32\pi^2-120\,\zeta(3)
 +236\right]+{\cal O}(\epsilon^2) \Big\}\,,
 \nn
\end{eqnarray}
which disagrees with $\tilde{{\cal A}}_1^{(2)}$ in terms proportional
to $\epsilon^m$ for $m>0$, and with $\tilde{{\cal A}}_2^{(2)}$ for all
terms in the $\epsilon$ expansion. 

\section{Spin Triplet Currents} 
\label{sectiontriplet}

In this section we discuss the form of the interpolating currents describing
the production of a fermion-antifermion pair in a spin triplet $S=1$ state for
arbitrary relative angular momentum $L$ (${}^{2S+1}L_J={}^3L_J$). As in
Sec.~\ref{sectionsinglet} the currents are defined in the c.m.\,frame.

\subsection*{Pauli Matrices}

Since the treatment of Pauli $\sigma$-matrices in $n$ dimensions involves a
number of subtleties, we briefly review some of their properties relevant for
the formulation of the currents. Many of the properties can be directly
obtained from the corresponding properties of the $\gamma$-matrices in $d=n+1$
dimensions. The 
$\sigma$-matrices $\sigma^i$ ($i=1,\ldots,n$) are the generators of SO(n)
rotations for spin $1/2$. They are traceless, hermitian, satisfy the Euclidean
Clifford algebra $\{\sigma^i,\sigma^j\}=2\delta^{ij}$, and can be defined
such that $\bmsigma^T=\bmsigma^*=-\sigma_2\bmsigma\sigma_2$. While traces of
a product of an even number of $\sigma$-matrices in $n$ dimensions can be
expressed as a 
multiple of $\mbox{Tr}[\bmone]$ using the anticommutator, traces of a product
of an odd number are identically zero and for the case of three
$\sigma$-matrices can require additional
rules to yield results that are consistent with the relations known from $n=3$. 
In this respect the product of three different $\sigma$-matrices can be
somewhat considered the three-dimensional
analog of $\gamma_5$ in four dimensions.\footnote{ 
The vanishing of traces of a product of an odd number of $\sigma$-matrices
in $n$ dimensions
is related to the simultaneous use of the cyclicity property of the trace
operators and the Euclidean Clifford algebra relations for all  
$\sigma$-matrices. Products of an odd number of $\sigma$-matrices,
however, do not arise from insertions of
potentials which we consider in this work. Such traces can, however, arise
e.g.\,when the spin-dependent radiation of ultrasoft gluons or annihilation
potentials need to be considered. Thus, in general, the cyclicity
property of the trace operation should only be used after
renormalization is completed - if the Euclidean Clifford algebra is applied
for all  $\sigma$-matrices. 
}
As for the case of the $\gamma$-matrices~\cite{DuganGrin} products of
$\sigma$-matrices in arbitrary number of dimensions cannot be reduced to a
finite basis, but represent an infinite set of independent structures. In
analogy to Ref.\,\cite{DuganGrin} it is convenient to define 
\begin{eqnarray}
\sigma^{i_1\cdots i_m} & \equiv &
\sigma^{[i_1}\sigma^{i_2}\cdots \sigma^{i_m]}
\,,\quad (m=1,2,\ldots)
\end{eqnarray}
for the averaged antisymmetrized product of $\sigma$-matrices, so e.g.\,
$\sigma^{ij}=1/2(\sigma^i\sigma^j-\sigma^j\sigma^i)$. It is straightforward to
derive the following useful relations:
\begin{eqnarray}
\sigma^{i_1\cdots i_m}\sigma^\ell & = &
(-1)^m\,\Big(\, \sigma^{\ell i_1\cdots i_m} \, + \,
\sum_{j=1}^{m}(-1)^j\,\delta^{\ell i_j}\,\sigma^{i_1\cdots \ihat_j\cdots i_m}\Big)
\,,
\label{sigma1}
\\
\sigma^\ell\sigma^{i_1\cdots i_m} & = &
\sigma^{\ell i_1\cdots i_m} \, + \,
\sum_{j=1}^{m}(-1)^{j+1}\,\delta^{\ell i_j}\,\sigma^{i_1\cdots \ihat_j\cdots i_m}
\,,
\label{sigma2}
\\
\left[ [\sigma^j,\sigma^\ell],\sigma^{i_1\ldots i_m} \right]
  & = & 4(-1)^m\, \sum_{k=1}^{m}\, (-1)^{k+1}
   \left( \delta^{j i_k}\sigma^{i_1\ldots \ihat_k\ldots  i_m \ell}   
  -  \delta^{\ell i_k}\sigma^{i_1\ldots \ihat_k\ldots  i_m j} \right)
\,,
\label{sigma3}
\\
\frac{1}{2} \left( \sigma^{j}\sigma^{i_1\ldots i_m}\sigma^{\ell}
+ \sigma^{\ell}\sigma^{i_1\ldots i_m}\sigma^{j} \right)
& = &
(-1)^m\,\delta^{j\ell}\,\sigma^{i_1\dots i_m} 
\nonumber \\ 
&& + \sum_{k=1}^{m}\, (-1)^{k+1}\left( \delta^{j i_k}\sigma^{i_1\ldots
  \ihat_k\ldots  i_m \ell} 
  +  \delta^{\ell i_k}\sigma^{i_1\dots \ihat_k\ldots  i_m j} \right)
\,,
\label{sigma4}
\\
\sum_k\,\sigma^k\sigma^{i_1\dots i_m}\sigma^k & = &
(-1)^m\,(n-2m)\,\sigma^{i_1\dots i_m}
\,,
\label{sigma5}
\\
\sum_{i_1\dots i_k}\,\sigma^{i_1\dots i_k}\,\sigma^{i_1\dots i_k}
& = &
(-1)^{\frac{k(k-1)}{2}}\,n\dots (n-k+1)
\, = \,
(-1)^{\frac{k(k-1)}{2}}\,\frac{\Gamma(n+1)}{\Gamma(n-k+1)}\,,\quad
\nn\\
\label{sigma6}
\\
\mbox{Tr}(\sigma^{i_1}\dots \sigma^{i_{2m}}) &=& \mbox{Tr}\,[\bmone]
 \!\!\! \!\!\!\sum_{\mathrm{inequ. \,\,permut.}\atop (p_1\dots p_{2m})\,\mathrm{of}\, (1\dots 2m) }
\!\!\!\!\!\!\!\!\!
\delta^{i_{p_1}i_{p_2}}\dots\delta^{i_{p_{2m-1}}i_{p_{2m}}}\,\delta_P\,.
\label{sigma7}
\end{eqnarray}
In the last relation, $\delta_P$ is the signature (sign) of the respective
permutation  
${1\;\,2 \,\ldots\, \;2m \choose p_1\,p_2\ldots \,p_{2m}}$, and for 
each $\delta^{i_m i_n}$ only $m<n$ is allowed. Note that we call two
permutations equivalent if they lead to the same term in the sum.
For $n=3$ we have $\sigma^{ij}=i \epsilon^{ijk}\sigma^k$, 
$\sigma^{ijk}=i \epsilon^{ijk}\bmone$, and $\sigma^{i_1\cdots i_m}=0$ for
$m>3$. Thus the latter are evanescent for $n\neq 3$. For $m\leq 3$ the
$\sigma^{i_1\cdots i_m}$ are related to physical spin operators.

\subsection*{S-Wave Currents}

The general interpolating current describing the production of a
fermion-antifermion pair in an S-wave state and in an arbitrary spin state in
$n$ dimensions has the form
\begin{eqnarray}
(j_{L=0})^{[i_1\cdots i_m]} & = &
\psi_{\bmp}^\dagger(x)\, \sigma^{i_1\cdots i_m}\,
(i\sigma_2)\chi_{-\bmp}^*(x)
\,.
\label{Swavecurrent1}
\end{eqnarray}
A SO(n) rotation by an angle $\theta$ around the axis $\bmn$ leads to the
transformed current 
$D_\theta(\psi_{\bmp}^\dagger\, \sigma^{i_1\cdots i_m}\,
(i\sigma_2)\chi_{-\bmp}^*)=\psi_{\bmp}^\dagger S(\theta)^\dagger \sigma^{i_1\cdots i_m}\,
S(\theta)(i\sigma_2)\chi_{-\bmp}^*$, where 
$S(\theta)=\exp(-i \theta \bmn.\bmsigma/2)$. Since 
$S(\theta)^\dagger \sigma^i S(\theta)=D^{ij}(\theta)\sigma^j$, where 
$D^{ij}(\theta)$ is the rotation matrix for $n$-vectors, the currents in
Eq.\,(\ref{Swavecurrent1}) are tensors with respect to SO(n) rotations. The
tensors are irreducible due to the antisymmetry of the 
$\sigma^{i_1\cdots i_m}$~\cite{grouptheory} and each have 
$(^n_m)$ independent free components. Their eigenvalues with respect to the
square of the total spin operator 
$\bmS^2=[\bmsigma_p\otimes\bmone_{ap}+\bmone_{p}\otimes\bmsigma_{ap}]^2/4=
(n[\bmone_a\otimes\bmone_{ap}]+[\bmsigma_p\otimes\bmsigma_{ap}])/2$, where
the indices 
of the $\sigma$-matrices are summed over, read
\begin{eqnarray}
\bmS^2\,(
j_{L=0})^{[i_1\cdots i_m]}
 & = &
\frac{1}{2}
\psi_{\bmp}^\dagger\,\Big( 
n\,\sigma^{i_1\cdots i_m}(i\sigma_2) + 
\bmsigma\sigma^{i_1\cdots i_m}(i\sigma_2)\bmsigma^T \Big)\,
\chi_{-\bmp}^*
\nn\\
& = &
\frac{1}{2}
\left(n+(-1)^m\,(2m-n)
\right)\,
(j_{L=0})^{[i_1\cdots i_m]}
\,.\qquad
\label{S2op}
\end{eqnarray}
For the physical currents with $m=(0,1,2,3)$ one thus finds the spin
eigenvalues $(0,n-1,2,n-3)$. Note that the spin eigenvalue for the unphysical
S-wave currents $\sigma^{i_1\cdots i_m}$ for $m>3$ are non-zero.
While for $n=3$ the spin singlet currents for
$m=0,3$ are equivalent, and likewise the triplet currents for $m=1,2$, each of
the currents represents a different irreducible representation of SO(n) for $n\neq
3$. It is a very instructive fact that the action of $\bmS^2$ onto the
singlet current $j_{L=0}^{[i_1i_2i_3]}$ does not give zero for $n\neq
3$. Thus to achieve that spin-dependent potentials do not contribute for
physical predictions involving spin singlet currents, in general additional finite
renormalizations are required in analogy to Ref.~\cite{DuganGrin}, unless a
scheme for potentials or currents is chosen such that they vanish
automatically. However, even if such a scheme is adopted, matrix elements,
matching conditions and anomalous dimensions can depend on the spin-dependent
potentials at nontrivial subleading order\footnote{
To be more specific,
we refer to orders of perturbation theory where subdivergences need to be
renormalized.
}.

All of the physical currents can arise in important processes. In
Tab.~\ref{tab1} the leading order matching for a number of
different currents is displayed. In general there are several
relativistic currents that can lead to the same nonrelativistic
current~\cite{Fadin:1991zw}. Note that the respective production
and annihilation currents are related either by hermitian or
antihermitian conjugation. Except for $\gamma_5$, which is
treated as fully anticommuting ($\gamma_5=(^{0\,\bmI}_{\bmI\,0})$),
three dimensional relations have not been used.
The full theory 
current with the $\gamma$-structure $\gamma^{ijk}$ for example arises for
fermion pair production in $\gamma\gamma$ collisions, as shown in 
Appendix~\ref{appendix2}.
However, also evanescent currents naturally occur in standard processes, such as
the current $j_{L=0}^{[i_1\cdots i_5]}$ that arises in fermion-antifermion
pair annihilation into three photons. The explicit computation can be found in
Appendix~\ref{appendix3}. Also note that the differences between
the two different singlet and triplet currents correspond to evanescent
operators as well. 
\begin{table}
\begin{center}
\begin{tabular}{|c||cc|}\hline  Full theory $\Gamma$ & &
$\Gamma(\bmp,\bm{\sigma})$  \\ \hline \hline 
 $\gamma^{i_1\ldots i_{2k}}$ && $(-1)^{k+1} \frac{p^\ell}{m}\,\sigma^{\ell
 i_1\ldots i_{2k}}$ \\ \hline 
 $\gamma^{i_1\ldots i_{2k+1}}$ && $(-1)^{k} \,\sigma^{ i_1\ldots i_{2k+1}}$
 \\ \hline 
 $\gamma^0\gamma^{i_1\ldots i_{2k}}$ &  $(\star)$ & $(-1)^{k} \sum_{j=1}^{2k}
 (-1)^{j+1}\frac{p^{i_j}}{m} \sigma^{ i_1\ldots\ihat_j\ldots i_{2k}}$ \\
 \hline 
 $\gamma^0\gamma^{i_1\ldots i_{2k+1}}$ & $(\star)$ & $(-1)^{k} \,\sigma^{ i_1\ldots
 i_{2k+1}}$  \\ \hline 
 $\gamma^{i_1\ldots i_{2k}}\gamma^5$ & $(\star)$ & $(-1)^{k} \,\sigma^{ i_1\ldots i_{2k}}$
 \\ \hline 
$\gamma^{i_1\ldots i_{2k+1}}\gamma^5$ & $(\star)$ & $(-1)^{k}\frac{p^{\ell}}{m}
 \sigma^{\ell i_1\dots i_{2k+1}}$ \\
 \hline 
$\gamma^0\gamma^{i_1\ldots i_{2k}}\gamma^5$ && $(-1)^{k} \,\sigma^{ i_1\ldots
   i_{2k}}$    \\ \hline 
$\gamma^0\gamma^{i_1\ldots i_{2k+1}}\gamma^5$ && $(-1)^{k} \sum_{j=1}^{2k+1}
 (-1)^{j}\frac{p^{i_j}}{m} \sigma^{ i_1\ldots\ihat_j\ldots i_{2k+1}}$    \\
 \hline 
 \end{tabular} 
\end{center}
\caption{Nonrelativistic production currents 
$\psi_{\bmp}^\dagger\,\Gamma(\bmp,\bm{\sigma}) \,
(i\sigma_2)\chi_{-\bmp}^*$ arising from the leading order matching with the full
theory currents $\bar{\psi}\,\Gamma \, \psi$. The nonrelativistic
normalization of the full theory spinors has been used (see
Eq.~(\ref{spinors})). The notation $\gamma^{i_1\ldots i_{m}}$ stands for the
averaged antisymmetrized product of $\gamma$-matrices for indices other than
zero. For the corresponding annihilation currents 
$\chi_{-\bmp}^T(-i\sigma_2)\,\Gamma^\prime(\bmp,\bm{\sigma})\,\psi_{\bmp}$ 
one has $\Gamma^\prime=-\Gamma$ for all entries 
with the $(\star)$ symbol and $\Gamma^\prime=\Gamma$ for the others.
}  
\label{tab1} 
\end{table}
It is well known from subleading order computations based on the effective
weak Hamiltonian that one needs to consistently account for the evanescent
operator structures that arise in matrix elements of physical operators when
being dressed with gluons. In Ref.\,\cite{DuganGrin} it was shown that a
renormalization scheme can be adopted such that a mixing of evanescent
operators into physical ones does not arise. Moreover it is also
known~\cite{HerrlichNierste,ChetyrkinMisiak} that modifications of the
evanescent operator basis (e.g.\,by adding physical operators multiplied by
functions of $\epsilon$ that vanish for $\epsilon\to 0$) and similar
modifications to the physical operator basis correspond to a change of the
renormalization scheme. While this does not affect physical predictions, it
does affect matrix elements, matching conditions and anomalous dimensions at
nontrivial subleading order.
Thus precise definitions of the schemes being employed have to be given to
render such intermediate results useful. 

In the framework of the nonrelativistic EFT these properties still apply.
However, the velocity power counting in the EFT allows for even
more specific statements. Concerning interactions through potentials,
transitions between S-wave currents in Eq.\,(\ref{Swavecurrent1}) that are
inequivalent cannot occur because the potentials are 
SO(n) scalars and the currents are (due to the different symmetry
patterns of their indices~\cite{grouptheory}) inequivalent irreducible
representations of SO(n). Even for currents with $L\neq 0$ and for the
spin-dependent spin-orbit and tensor potentials (which is all we need
to consider at NNLL order) one can show with
Eqs.\,(\ref{sigma3}) and (\ref{sigma4}) that transitions between currents
containing $\sigma^{i_1\cdots i_m}$ with a different number of indices
cannot occur (see Sec.\,\ref{sectionNLLtriplet}.1). Transitions can,
however, arise between currents with 
different angular momentum, such as for the tensor potential that can
generate transitions between $L$ and $L^\prime=L\pm 2$ (see
Sec.\,\ref{sectionNLLtriplet}). The same arguments apply to the
exchange of soft gluons (in the framework of vNRQCD) since they cannot appear
as external particles and furthermore need to be exchanged in pairs due to
energy conservation. In this respect the effects from soft gluon exchange
effectively represent modifications to the potential interactions (see
e.g. Ref.\,\cite{LMR}). Concerning the 
exchange of ultrasoft gluons, transitions between currents
containing $\sigma^{i_1\cdots i_m}$ with a different number of indices
can arise, but only if the interaction is spin-dependent. The dominant
among these interactions corresponds to the operator 
$\psi_{\bmp}^\dagger\sigma^{ij} k^j \psi_{\bmp} A^i$, where $A^i$ is
an ultrasoft gluon. This operator can only
contribute at N${}^4$LL order, which is beyond the present need and
technical capabilities. Thus in practice, transitions between
currents containing $\sigma^{i_1\cdots i_m}$ with a different number
of indices do not need to be considered. 

So for the S-wave currents in $n$ dimensions one can employ either one 
of the two spin singlet ($k=0,3$) or triplet currents ($k=1,2$) in the
EFT and the difference corresponds to a change in the renormalization
scheme. This means in particular that as long as the renormalization
process is restricted e.g.\,to time-ordered products of the currents, one
can (but does not have to) freely use three-dimensional relations to reduce
the basis of the physical currents. However, once the basis of the
physical currents is fixed (where each current is irreducible with
respect to SO(n)), one has to consistently apply the computational
rules in $n$  dimensions discussed in the previous sections. As an
example, instead of using the current 
$\psi_{\bmp}^\dagger\sigma^{ijk}(i\sigma_2)\chi_{-\bmp}^*$ that arises
in $\gamma\gamma$ collision, one can employ the current 
$\psi_{\bmp}^\dagger(i\sigma_2)\chi_{-\bmp}^*$, defined in $n$ dimensions,
times the $\epsilon$-tensor $i\epsilon^{ijk}$ defined in three dimensions,
which means that the $\epsilon$-tensor is zero if any of its indices takes a
value different from $1,2,3$. Concerning the
$\epsilon$-tensor, this works because the $\epsilon$-tensor does not
play any role during 
the renormalization procedure as long as we only consider time-ordered
products of the currents.\footnote{This procedure cannot be applied in
this simple form, if the initial state that is involved in the
quark-antiquark production process is also involved in the
renormalization procedure, as can be the case for QED corrections.} 
This justifies the approach in Sec.\,\ref{sectionsinglet} where we
have only considered spin singlet currents for fermions involving the
current $\psi_{\bmp}^\dagger(i\sigma_2)\chi_{-\bmp}^*$. This scheme is
also advantageous practically because for the current
$\psi_{\bmp}^\dagger\sigma^{ijk}(i\sigma_2)\chi_{-\bmp}^*$ the
matrix elements of the spin-dependent potentials that vanish for $n=3$
can, as  discussed above, give evanescent contributions for $n\ne 3$. 
Moreover, from these considerations we can also conclude that currents
containing evanescent $\sigma^{i_1\cdots i_m}$ matrices ($m>3$) can
(but do not have to) be dropped from the very beginning in the EFT as
long as one does not need to account for spin-dependent ultrasoft gluon
interactions.  

An important lesson to learn from this discussion is that partial results at
nontrivial subleading order obtained from the threshold 
expansion~\cite{Beneke} of full theory diagrams such as e.g. the 
contributions from the hard regions obtained in
Refs.~\cite{Czarnecki1,Beneke4,firstc0,2loop_hard}, 
are equal to the EFT matching conditions only in schemes for the effective 
theory currents and potentials that are compatible with the nonrelativistic
reduction of the $\gamma$ matrices that has been used during the threshold 
expansion. In general, there are additional contributions to EFT matching
conditions to account for the scheme choices made in the EFT.
Note that in the threshold expansion different scheme choices can also 
be possible, in particular for the treatment of $\gamma_5$.
In Sec.~\ref{sectioncomments} we comment on scheme dependences
of a number of partial results that can be found in the literature.

At this point one might also ask whether $\gamma_5$ needs special
treatment in the nonrelativistic EFT in $n$ dimensions. Chirality
and the flavor symmetries are not relevant in the EFT for a single heavy
particle-antiparticle system and their potential
effects are contained in the matching contributions of the EFT to 
the full theory. For the
matching relations shown in Tab.~\ref{tab1} we have used a totally
anticommuting $\gamma_5$. Since the resulting effective theory
currents have well defined SO(n) transformation properties and reduce
to the proper nonrelativistic currents for $n\to 3$, this represents a
consistent scheme choice from the point of view of the effective theory
computations. A different ansatz for $\gamma_5$ such as
$\gamma_5=i\gamma^0\gamma^1\gamma^2\gamma^3$ (which is the fully consistent
one in the full theory) just corresponds to a different choice of scheme for  
the currents in the nonrelativistic EFT, and both schemes can be used
consistently. Note that from the results in Tab.~\ref{tab1} without an
explicit $\gamma_5$ one can also 
derive the form of the EFT currents for the ansatz
$\gamma_5=i\gamma^0\gamma^1\gamma^2\gamma^3$ (see also
Sec.~\ref{sectioncomments}).

\subsection*{Arbitrary Angular Momentum}

Based on the results obtained in the previous sections it is
straightforward to construct spin-triplet currents with arbitrary
angular momentum $L$ (${}^{2S+1}L_J={}^3L_J$)). They can be obtained
by determining irreducible SO(n) representations from products of the
tensors $T^{i_1\dots i_L}(\bmp)$ describing angular momentum $L$ and
the spin-triplet $S=1$ currents discussed in the previous
paragraphs. Due to the different symmetry patterns of the symmetric
$T^{i_1\dots i_L}(\bmp)$ tensors and the antisymmetric 
$\psi_{\bmp}^\dagger\sigma^{i_1\cdots i_k}(i\sigma_2)\chi_{-\bmp}^*$
currents one needs to apply the construction principles for general
tensors~\cite{grouptheory} which state that tensors are irreducible
with respect to SO(n) if and only if they are traceless and if their
indices have a symmetry pattern according to a standard Young tableau. 
As for the case of the S-wave currents the physical basis for
arbitrary spatial angular momentum is not unique due to the existence
of evanescent operator structures. 

Here we construct currents with fully symmetric indices because of
their comparatively simple form and because their number of indices is
equal to the total angular momentum $J$ quantum number. For the
case $J=L\pm 1$, these currents are contained in the reduction of the 
reducible tensor $A^{i_1\cdots i_{L+1}} = \psi_{\bmp}^\dagger 
\,T^{i_1\dots i_L}(\bmp)\sigma^{i_{L+1}}(i\sigma_2)\chi_{-\bmp}^*$. Upon
symmetrization and removal of all traces one obtains the 
${}^{2S+1}L_J={}^3L_{L+1}$ current
\begin{eqnarray}
 \left(j_{J=L+1}^{S=1}\right)^{i_1\dots i_{L+1}} 
 & \equiv &  
\psi_{\bmp}^\dagger
\,\Big[\Gamma^{S=1}_{L+1}(\bmp,\bm{\sigma})^{i_1\dots i_{L+1}} \Big]
(i\sigma_2)\chi_{-\bmp}^* 
\,, \nn\\
 \Gamma^{S=1}_{L+1}(\bmp,\bm{\sigma})^{i_1\dots i_{L+1}}  
&\equiv& 
\sum_{k=1}^{L+1}\,T^{\,i_1\dots\ihat_k\dots i_{L+1}}(\bmp)\,\sigma^{i_k}
  - \frac{2}{2L+n-2}\,\sum_{k<j \atop k=1}^{L+1}\,\delta^{i_k
  i_j}\,T^{\,i_1\dots\ihat_k\dots\ihat_j \dots i_{L+1}
  \ell}(\bmp)\,\sigma^{\ell}
\,.\nn\\
\label{tripletcurrent1}
\end{eqnarray}
The subtractions in the second term of Eq.~(\ref{tripletcurrent1})
needed to achieve zero traces are for themselves
${}^{2S+1}L_J={}^3L_{L-1}$ currents, so we can define
\begin{eqnarray}
 \left(j_{J=L-1}^{S=1}\right)^{i_1\dots i_{L-1}} 
&\equiv &
 \psi_{\bmp}^\dagger
 \Big[\,\Gamma^{S=1}_{L-1}(\bmp,\bm{\sigma})^{i_1\dots i_{L-1}}\Big]
(i\sigma_2)\chi_{-\bmp}^* 
\,,\nn\\[2mm]
\Gamma^{S=1}_{L-1}(\bmp,\bm{\sigma})^{i_1\dots i_{L-1}}
  & \equiv & 
T^{\,i_1\dots i_{L-1}\ell}(\bmp)\,\sigma^{\ell}
\,.
\label{tripletcurrent2}
\end{eqnarray}
The tensor  $A^{i_1\cdots i_{L+1}}$ also contains ${}^3L_{L}$
currents. However, their indices are not fully symmetric. Their form
is discussed in the Appendix~\ref{appendix3} together with a more
detailed derivation of the two currents displayed above. A fully
symmetric ${}^3L_{L}$ current can be obtained from the reduction of
the tensor  $B^{i_1\dots i_{L+2}} = \psi_{\bmp}^\dagger 
\,T^{i_1\dots i_L}(\bmp)\sigma^{i_{L+1}
  i_{L+2}}(i\sigma_2)\chi_{-\bmp}^*$. Upon 
removal of traces of  $B^{i_1\dots i_{L+2}}$ one
can identify the  ${}^3L_{L}$ current
\begin{eqnarray}
 \left(j_{J=L}^{S=1}\right)^{i_1\dots i_{L}}  
 &\equiv &
 \psi_{\bmp}^\dagger
 \Big[\,\Gamma^{S=1}_{L}(\bmp,\bm{\sigma})^{i_1\dots i_{L}}\Big]
(i\sigma_2)\chi_{-\bmp}^* 
\,,\nn\\
\Gamma^{S=1}_{L}(\bmp,\bm{\sigma})^{i_1\dots i_{L}}
 & \equiv & 
\sum_{k=1}^{L}\,T^{\,i_1\dots\ihat_k\dots i_{L} \ell}(\bmp) \,\sigma^{\ell\,
 i_k}  
\,.
\label{tripletcurrent3}
\end{eqnarray}
The tensor $B^{i_1\dots i_{L+2}}$ also contains evanescent currents and also
currents describing  ${}^3L_{L\pm 1}$ states that differ from the currents
given above for $n\neq 3$. Their form is also discussed in the
Appendix~\ref{appendix3}. The number of independent components of the
spin-triplet currents defined above is that of a fully symmetric tensor with
$J$ indices, {\it i.e.} equal to $n_J$ given in Eq.~(\ref{nL}). All three
currents have parity and charge conjugation $P=(-1)^{L+1}$ and $C=(-1)^{L+S}$,
respectively. Also note that that 
$\Gamma^{S=1}_{L\pm 1}(\bmp,\bm{\sigma})$ and 
$i\Gamma^{S=1}_{L}(\bmp,\bm{\sigma})$ are hermitian matrices.

The eigenvalue equation of the currents with respect to the spin-orbit
operator $\bmS.\bmL= - [S^j,S^k] \,p^j \partial/ \partial p^k 
= - 1/4( \, [\bm{\sigma}_p^j,\bm{\sigma}_p^k]\otimes\bmone_{ap} +
\bmone_p\otimes[\bm{\sigma}_{ap}^j,\bm{\sigma}_{ap}^k] \, )\,p^j \partial/
\partial p^k $ has the form  
\begin{eqnarray}
\lefteqn{
(\bmS.\bmL)\,\Big(
\psi_{\bmp}^\dagger \, \Gamma(\bmp,\bmsigma)^{i_1\cdots i_J}(i\sigma_2)\,\chi_{-\bmp}^*\Big) 
 = } \nn\\
&& =
-\frac{1}{4}\, \psi_{\bmp}^\dagger\Big( [\bm{\sigma}^j,\bm{\sigma}^k]
\,p^j \frac{\partial}{\partial p^k} \,\Gamma(\bmp,\bm{\sigma})^{i_1\cdots
  i_J}(i\sigma_2)+ 
\,p^j \frac{\partial}{\partial p^k} \,\Gamma(\bmp,\bm{\sigma})^{i_1\cdots
  i_J}(i\sigma_2)[ 
\bm{\sigma}^j,\bm{\sigma}^k]^T \Big) \chi_{-\bmp}^*
\,,\mbox{\quad}
\label{SLop}
\end{eqnarray}
where $\Gamma(\bmp,\bmsigma)^{i_1\cdots i_J}$ represents any of the
tensors between the fermion fields.
The eigenvalues of the different currents defined in 
Eqs.~(\ref{singletcurrent1},\ref{tripletcurrent1},\ref{tripletcurrent2},\ref{tripletcurrent3})   
with respect to the operators $\bmL^2$,
$\bmS^2$, $\bmL.\bmS$ and $\bmJ^2$ are summarized in Tab.~\ref{tab2}. 
\begin{table}
\begin{center}
\begin{tabular}{|c||c|c|c|c|}\hline     
& \raisebox{-1mm}[1mm]{$j_{J=L}^{S=0}$} 
& \raisebox{-1mm}[1mm]{$j_{J=L-1}^{S=1}$} 
& \raisebox{-1mm}[1mm]{$j_{J=L}^{S=1}$} 
& \raisebox{-1mm}[1mm]{$j_{J=L+1}^{S=1}$}
  \\[2mm] \hline \hline  
$\bmL^2$  & $L(L+n-2)$ & $L(L+n-2)$ & $L(L+n-2)$ & $L(L+n-2)$ \\[2mm] \hline 
$\bmS^2$ & 0 & $n-1$ & 2 & $n-1$   \\[2mm] \hline 
$\bmS.\bmL$ & 0 & $-(L+n-2)$ & $-(n-2)$ & $L$ \\ \hline 
$\bmJ^2$ &  $L(L+n-2)$ & $J(J+n-2)$ & $J(J+n-2)+6-2n$ & $J(J+n-2)$\\ \hline  
 \end{tabular} 
\end{center}
\caption{
Spin and orbital angular momentum quantum numbers of the currents defined in
Eqs.~(\ref{singletcurrent1},\ref{tripletcurrent1},\ref{tripletcurrent2},\ref{tripletcurrent3})
in $n$ dimensions.}
\label{tab2} 
\end{table}
For the computation of the NLL order anomalous dimensions of the
currents from time-ordered products of two currents
(Secs.~\ref{sectionNLLsinglet} and \ref{sectionNLLtriplet}), the total 
contractions of spatial and spin indices (i.e. taking the trace of
products of $\sigma$-matrices) are useful:
\begin{eqnarray}
\mbox{Tr}\Big[\Gamma^{S=1\,\dagger}_{L+1}(\bmp,\bmsigma)^{i_1\cdots i_{L+1}}\,
\Gamma^{S=1}_{L+1}(\bmq,\bmsigma)^{i_1\cdots i_{L+1}}\Big] 
& = & 
(L+1)(2L+n)\,\frac{L+n-2}{2L+n-2}\,T_{L,n}(\bmp,\bmq)\,
\mbox{Tr}\,[\bmone]
\,,
\nn\\[2mm]
\mbox{Tr}\Big[\Gamma^{S=1\,\dagger}_{L}(\bmp,\bmsigma)^{i_1\cdots i_{L}}\,
\Gamma^{S=1}_{L}(\bmq,\bmsigma)^{i_1\cdots i_{L}}\Big] 
& = & 
L(L+n-2)\,T_{L,n}(\bmp,\bmq)  \, \mbox{Tr}\,[\bmone]
\,,
\nn\\[2mm]
\mbox{Tr}\Big[\Gamma^{S=1\,\dagger}_{L-1}(\bmp,\bmsigma)^{i_1\cdots i_{L-1}}\,
\Gamma^{S=1}_{L-1}(\bmq,\bmsigma)^{i_1\cdots i_{L-1}}\Big] 
& = & 
T_{L,n}(\bmp,\bmq) \, \mbox{Tr}\,[\bmone]
\,,
\end{eqnarray}
where the function $T_{L,n}$ is defined in Eq.~(\ref{Tcontract}).
Note that the expressions only involve traces of products of an even
number of $\sigma$-matrices which can be computed unambiguously from
the anticommutator of two $\sigma$-matrices.

\section{NLL Order anomalous dimensions: Singlet Case} 
\label{sectionNLLsinglet}

For the computation of the renormalization constants of the physical
currents defined in the previous sections one has to determine the
overall UV-divergence of particle-antiparticle-to-vacuum on-shell
matrix elements of the currents from vertex diagrams. However, this
method is complicated by the fact that for on-shell external particles
with $E=\bmp^2/m$ the vertex diagrams also contain IR-divergent
Coulomb phases that have to be distinguished from the
UV-divergences. To avoid this issue we use the method from
Ref.~\cite{Hoang3loop} which considers current-current correlator
graphs obtained from closing the external particles-antiparticle line
of the vertex diagrams with an additional insertion of the (hermitian
conjugated) current. This means that one has to determine diagrams with
one more loop, but in this way all IR-divergences associated to the
Coulomb phase cancel and one needs to compute a fewer number of
diagrams. Technically, the UV-divergences associated to the currents
appear as subdivergences of the correlator diagrams, which can,
however, easily identified from the absorptive parts of the
correlators. 

\begin{figure}[t] 
\begin{center}
 \leavevmode
 \epsfxsize=14cm
 \leavevmode
 \epsffile[45 460 545 575]{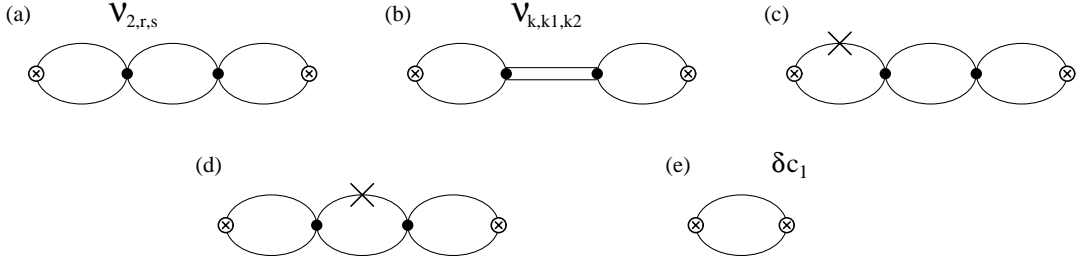}
 \vskip  0.0cm
 \caption{
Three-loop current correlator diagrams and the counterterm diagram for
the computation of the NLL anomalous dimension of the
currents. Combinatorial factors are suppressed.
 \label{fignllcorrelators} }
\end{center}
\end{figure}

Since there are no UV divergences in the EFT that can lead to a LL
anomalous dimension of the currents that are leading in the
nonrelativistic expansion~\cite{LMR}, the NLL order level represents
the lowest order in which these currents are renormalized.
The corresponding three-loop current correlator diagrams
are depicted in Fig.\,\ref{fignllcorrelators}.
%
%
The terms
${\cal V}_c^{(s)}$ and ${\cal V}_2^{(s)}$, ${\cal V}_r^{(s)}$ are the
color singlet Wilson coefficients of the spin-independent potentials of 
order $\alpha_s v^{-1}$ ($1/\bmk^2$) and $\alpha_s v$ ($1/m^2$,
$(\bmp^2+\bmp^{\prime 2})/(2m^2\bmk^2)$),
respectively~\cite{amis,amis2}, while ${\cal V}_{k,k1,k2}^{(s)}$ are 
the Color singlet Wilson coefficients of the sum operators 
${\cal O}_{k}^{(s)}$, ${\cal O}_{k1}^{(1)}$, ${\cal O}_{k2}^{(T)}$ 
introduced in Refs.\,\cite{HoangStewartultra,Hoang3loop}. The latter operators
are the analogue of the $1/m|\bmk|$ non-Abelian-potential that was used in
early computations where no systematic renormalization group
improvement was carried out (see e.g.~\cite{HT1}). We note that the
Wilson coefficient  ${\cal V}_2^{(s)}$ differs for quarks and for
scalars~\cite{HoangRuizSvNRQCD}. This is, however, irrelevant for the result
of the anomalous dimension and only affects the solution of the resulting
renormalization group equations (RGE's). 
The diagram (e) contains the renormalization constant of the current
which we determine in the $\msb$ scheme. Generically we define the
renormalization constant as
\begin{eqnarray}
Z_c & = & 1 + \frac{\delta z_c^{\rm NLL}}{\epsilon}\,+\,\ldots
\,.
\end{eqnarray}
The resulting NLL order RGE for the current
Wilson coefficients $c(\nu)$ reads 
\begin{eqnarray}
\nu\frac{\partial}{\partial\nu}\ln[c(\nu)] & = &
4\,\delta z_c^{\rm NLL}
\,.
\label{rgeCurrent}
\end{eqnarray}

Due to rotational invariance and the fact that the currents are
irreducible, all components of any given current have the same
anomalous dimension. Therefore we can compute the correlators for
all spatial and spin indices being contracted. As an example, for the
correlators in Fig.~\ref{fignllcorrelators}a with the two different
insertions of the 
potential $-i{\cal V}_2/m^2$, and after applying the contraction formula
(\ref{Tcontract}), one has to compute the three-loop diagram 
\begin{eqnarray}
-2i m N_c\,{\cal V}_c^{(s)}\,{\cal V}_2^{(s)}\,N_L
\int\!\! D^n \bmq_1\, D^n \bmq_2\, D^n \bmq_3 \,
\frac{|\bmq_1|^L|\bmq_3|^L P_L(n,\bar\bmq_1.\bar\bmq_3)}
{(\bmq_1^2+\delta)\,(\bmq_2^2+\delta)\,(\bmq_2-\bmq_3)^2\,(\bmq_3^2+\delta)}
\,.
\label{V2correlatorexample}
\end{eqnarray}
The expression in Eq.~(\ref{V2correlatorexample}) is equivalent to the
corresponding S-wave result up to the weight factor 
$N_L|\bmq_1|^L |\bmq_3|^L P_L(n,\bar\bmq_1.\bar\bmq_3)$, where
$\bmq_1$ and $\bmq_3$ are the loop-momenta on the sides of the
diagram. The weight factor is a homogeneous polynomial of degree $L$
of the terms $\bmq_1^2$, $\bmq_3^2$ and $\bmq_1.\bmq_3$, see
Eq.~(\ref{Legendredef}). The 
contributions to the renormalization constant $Z_c$ of each of the
diagrams in Fig.~\ref{fignllcorrelators} with the weight factor
$(\bmq_1.\bmq_3)^m(\bmq_1^2)^{\frac{L-m}{2}}(\bmq_3^2)^{\frac{L-m}{2}}$
and proper combinatorial 
factors are as follows:
\begin{eqnarray}
\delta z^{\rm NLL}_{c,(a)} & = & 
-\frac{{\cal V}_c^{(s)}(\nu){\cal V}_r^{(s)}(\nu)}
   {64\pi^2}\,
   {}_2F_1({\textstyle\frac{1}{2},-m,\frac{3}{2},2}) 
\, + \,
\left\{
\begin{array}{c}
0\,,\qquad m\,\,\mbox{odd}\\[2mm]
{\displaystyle -\frac{{\cal V}_c^{(s)}(\nu){\cal V}_2^{(s)}(\nu)}{64(1+n)\pi^2}}
\,,m\,\,\mbox{even}
\end{array}
\right.
\,,
\nn\\[2mm]
\delta z^{\rm NLL}_{c,(b)} & = &
\Big(
\frac{\alpha_s(m\nu)^2}{4} \big(3{\cal V}_{k1}^{(s)}(\nu)+2{\cal V}_{k2}^{(s)}(\nu) \big)
+\frac{\alpha_s^2(m\nu)}{8}\big(\frac{C_F^2}{2}-C_AC_F\big)
\bigg)\,
   {}_2F_1({\textstyle\frac{1}{2},-m,\frac{3}{2},2})
\,,
\nn\\[2mm]
\delta z^{\rm NLL}_{c,(c,d)} & = &
-\frac{\big({\cal V}_c^{(s)}(\nu)\big)^2}
   {256\pi^2}\,
   {}_2F_1({\textstyle\frac{1}{2},-m,\frac{3}{2},2}) 
\,.
\end{eqnarray}
In App.~\ref{appendix5} some useful formulae to determine these
results are given.
The final result for the anomalous dimension can then be obtained by
using Eq.~(\ref{Legendredef}) for the definition of the generalized Legendre
polynomial. The result is remarkably simple and reads
\begin{eqnarray}
\delta z^{\rm NLL}_c & = & 
\frac{1}{4(2L+1)}\,\bigg[
-\frac{1}{16\pi^2}\,{\cal V}_c^{(s)}(\nu)\,
\Big( \frac{{\cal V}_c^{(s)}(\nu)}{4} + {\cal V}_r^{(s)}(\nu) \Big)
\nonumber\\[2mm] & & \qquad  +\,
\alpha_s(m\nu)^2 \Big(3{\cal V}_{k1}^{(s)}(\nu)+2{\cal V}_{k2}^{(s)}(\nu) \Big)
+\frac{\alpha_s^2(m\nu)}{2}\big(\frac{C_F^2}{2}-C_AC_F\big)
\,\bigg]
\nonumber\\[2mm] & & 
-\, \frac{{\cal V}_c^{(s)}(\nu){\cal V}_2^{(s)}(\nu)}{64\pi^2}\,\delta_{L0} 
\,.
\label{Zscalarpot}
\end{eqnarray}
The explicit expressions for the potential Wilson coefficients for quarks and
colored scalars can be found in Refs.~\cite{Hoang3loop} and
\cite{HoangRuizSvNRQCD} (see also Ref.~\cite{Pineda1}).
For S- and P-waves the expression agrees with previous results, see
e.g.~Ref.~\cite{HoangRuizSvNRQCD}. Written in this way, the 
result~(\ref{Zscalarpot}) is valid for  
both scalar and fermion singlet currents. Differences in the running 
of the singlet currents for scalars and fermions can only arise at this order
from a different running of the Wilson coefficients for the potentials.  
Since at this order only the coefficient ${\cal V}_2^{(s)}$ differs for quarks
and for scalars, and since ${\cal V}_2^{(s)}$ only contributes for S-waves, 
the NLL order anomalous dimensions for quarks and scalars agree for $L\ge 2$.
We note that the NLL anomalous dimension is
independent of the scheme used for the currents (or the potentials)
since at LL order the currents are not renormalized. Had we
e.g.\,used the singlet current with $\sigma^{ijk}$ the spin-dependent
potentials could in general contribute to the UV-finite terms at this order,
but not to the NLL anomalous dimension.

\section{NLL Order anomalous dimensions: Triplet Case} 
\label{sectionNLLtriplet}

\subsection*{Spin Dependent Potentials}

Before determining the NLL order anomalous dimension for
the spin triplet currents, it is instructive to briefly discuss the form of the
spin-dependent potentials from the point of view of scheme-dependences in the
EFT in $n$ dimensions - despite the fact that the results for the NLL order
anomalous dimensions of the currents defined above are
scheme-independent. From the expansion of full theory 
quark-antiquark scattering diagrams  without relying on any $\sigma$-matrix
relation for $n=3$ at intermediate steps, one obtains the tree-level potential  
($\bmk=\bmp^\prime-\bmp$)
\begin{eqnarray}
 V({\bmp},{\bmp^\prime}) &  = & 
g_s^2 (T^A \otimes \bar T^A) \bigg[\!
 \frac{1}{\bmk^2}
 + \frac{{\bmp^2 + \bmp^{\prime 2}}}{2 m^2 \bmk^2} - \frac{1}{2m^2 } 
 + \frac{k^l k^m}{m^2\bmk^2}
   \left[S_p^i,S_p^l\right]\left[S_{ap}^i,S_{ap}^m\right] 
 - \frac{3\,p^ip^{\prime j}}{2m^2\bmk^2}\left[S^i,S^j\right]
\!\! \bigg]
\nn\\[2mm] & = &
g_s^2 (T^A \otimes \bar T^A)\bigg[
 \frac{1}{\bmk^2}
 + \frac{{\bmp^2 + \bmp^{\prime 2}}}{2 m^2 \bmk^2} + \frac{n-3}{4m^2 } 
\nn\\[2mm] & & \qquad \qquad
+ \frac{\bmS^2}{m^2 } \left(\frac{\bmS^2}{n}-1 \right)
- \frac{3}{4 m^2 \bmk^2} \, S^{[ij]} \cdot T^{[p^i {p'}^j]}
+ \frac{2\bmS^2-n}{2 m^2 \bmk^2} \, S^{ij} \cdot T^{ij}(\bmk)
\, \bigg] \,,
\label{ndimpot}
\end{eqnarray}
where in Eqs.~(\ref{ndimpot}) and in the following the symbol for 
the particle-antiparticle tensor product is indicated only in color
space. The spin-dependent terms, displayed in the last equality, have been written
in terms of irreducible tensors with
\begin{eqnarray}
T^{[p^i {p'}^j]} &=& p^ip^{\prime\,j}-p^jp^{\prime\,i}\,,\nn\\
S^{[ij]}=[S^i,S^j] &,&
S^{ij}=\frac{1}{2}\{S^i,S^j\}-\frac{\delta^{ij}}{n}\bmS^2 \,.
\end{eqnarray}
The tree-level potential in Eq.~(\ref{ndimpot}) is consistent with the
$n$-dimensional expression used in Ref.~\cite{beneke} accounting for the fact that
an insertion of the last two terms in the second line of Eq.~(\ref{ndimpot})
does not contribute for the current
$j^{S=1}_{J=L+1=1}=\psi_{\bmp}^\dagger\, \sigma^{i}\,(i\sigma_2)\chi_{-\bmp}^*$  
at NNLO.\footnote{
While any number of insertions of the spin-orbit potential
$\sim S^{[ij]} T^{[p^i {p'}^j]}$ in the correlator of two
$j^{S=1}_{J=L+1=1}$ currents vanishes identically in $n$ dimensions, several
insertions of the tensor potential $\sim  S^{ij} T^{ij}(\bmk)$
can yield non-vanishing contributions beyond NNLO.}

Apart from the spin-orbit potential $\sim [S^i,S^j]p^i {p^\prime}^j$ which 
trivially reduces to $\sim \bmS.(\bmp\times\bmp^\prime)$ for $n=3$, the result 
does not have a very close resemblance to the well known spin-dependent terms
in three dimensions~\cite{TitardYndurain},
\begin{eqnarray}
 V^\prime({\bmp},{\bmp^\prime}) &  = & 
g_s^2 (T^A \otimes \bar T^A)\! \bigg[ -\frac{\bmS^2}{3 m^2 }
-\frac{3}{2m^2}\Lambda({\bmp^\prime},{\bmp})-\frac{1}{12m^2} \, T({\bmk}) \, \bigg] 
\,,
\label{spinpotdef1}
\end{eqnarray}
where 
\begin{equation}
{\bmS} \, = \, \frac{ {{\bmsigma}_p + {\bmsigma}_{ap}} }{2}\,,
 \quad 
 \Lambda({\bmp^\prime},{\bmp}) \, = \, -i \frac{{\bmS}( {\bmp^\prime}
 \times {\bmp}) }{  {\bmk}^2 }\,,\quad
 T({\bmk}) \, = \, {({\bmsigma}_{p}{\bmsigma}_{ap}}) - 
\frac{3\, { ({\bmsigma}_p\bmk})\,({\bmsigma}_{ap}\bmk)}{ {\bmk}^2} \,.
\end{equation}
This is because the three-dimensional relation 
$\sigma_{a}^{ij}\otimes \sigma_{ap}^{ik}=
\sigma_{a}^k\otimes\sigma_{ap}^j-
\delta^{jk}\sigma_{a}^l\otimes\sigma_{ap}^l$ 
does not hold in this form for $n\neq 3$. As for the  currents we are,
however, free to use any scheme for the spin-dependent potentials as long as
they are SO(n)-scalars and reduce to the known spin-dependent potentials for
$n=3$. 

A suitable scheme choice can be obtained from the $\sigma$ structures
shown in Eq.~(\ref{ndimpot}), or from three-dimensional potentials
in Eq.\,(\ref{spinpotdef1}) through the $n$-dimensional replacements
\begin{eqnarray}
 -i\bmS(\bmp'\times\bmp)
  &\longrightarrow &
 [S^i,S^j]\,p^i {p'}^j 
\, = \, 
\frac{1}{2}\, S^{[ij]} \cdot T^{[p^i {p'}^j]}
\,,
\nn\\[3mm]
 (\bmsigma_p\bmsigma_{ap})-3\,\frac{(\bmsigma_p\bmk)\,(\bmsigma_{ap}\bmk)}
  {\bmk^2} 
 &\longrightarrow &
 (\bmsigma_p\bmsigma_{ap})-n\,\frac{(\bmsigma_p\bmk)\,(\bmsigma_{ap}\bmk)}
 {k^2}
\, = \,
-2n\, S^{ij} \frac{T^{ij}(\bmk)}{\bmk^2}
\,.
\label{tensordef1}
\end{eqnarray} 
We will use the scheme shown in Eq.~(\ref{tensordef1}) for the
computations of the anomalous dimensions 
in the following sections. This scheme is very convenient because the
different potentials are 
total contractions of spin- and angular momentum tensors containing only one
single irreducible representation. As a comparison, consider in
contrast the scheme choice $(\bmsigma_p\bmsigma_{ap}) - 3
(\bmsigma_p\bmk)(\bmsigma_{ap}\bmk)/\bmk^2$ for the 
tensor potential in $n$ dimensions, which is in principle a viable
scheme as well since it is a scalar and also reduces to the correct
expression for $n=3$. Written in terms of irreducible tensors one
finds 
\begin{eqnarray}
{{\bmsigma}_p {\bmsigma}_{ap}} - 3\frac{({\bmsigma}_p\bmk)\, 
 ({\bmsigma}_{ap}\bmk) }{ {\bmk}^2}
& = &
-6\, S^{ij} \frac{T^{ij}(\bmk)}{\bmk^2} 
\, + \,
\frac{2(n-3)}{n}\,\bmS^2 - (n-3)
\,.
\label{tensordef2}
\end{eqnarray}
While the tensor potential based on Eq.~(\ref{tensordef1}) is a natural
minimal extension of the tensor potential in three dimensions, the tensor
potential based on Eq.~(\ref{tensordef2}) does implicitly lead to a change of
scheme for the $\bmS^2/m^2$ potential and the spin-independent $1/m^2$
potential, which in general affects matrix elements, matching conditions and
anomalous dimensions at non-trivial subleading order, {\it i.e.} beyond NLL
order. 

We would like to point out that the spin-dependent potentials in
Eqs.~(\ref{tensordef1}) cannot produce transitions between the spin triplet
currents with fully symmetric indices introduced in Sec.~\ref{sectiontriplet}
and other currents that transform as inequivalent irreducible representations 
of SO(n) for $n\neq 3$, but that have the same quantum numbers for $n=3$ 
(e.g. built from the tensors $\widetilde{\Gamma}^{S=1}_J$ introduced in the
Appendix.~\ref{appendix3}). Technically this property
arises because a given current with fully symmetric indices and quantum numbers
$J$ and $L$ (${\Gamma}^{S=1}_J$) 
always differs from its counterpart $\widetilde{\Gamma}^{S=1}_J$
in the number of sigma matrices. For example, the insertion of a spin-orbit
potential in a current with a spin structure $\sigma^{i_1\ldots i_m}$  yields
the spin tensor on the LHS of Eq.~(\ref{sigma3}), where the indices $j$ and
$\ell$ are contracted with the initial and final momenta. The RHS of the
same equation shows that the resulting spin structure consists of terms with
the same number of sigma matrices as the initial current. The same holds
trivially for the case of the $\bmS^2$ potential, and also for the tensor
potential, where the relevant relation is Eq.~(\ref{sigma4}). The possibility
that the spin-dependent $\bmS^2$ and the spin-orbit potentials could induce
transitions between currents 
transforming according to different irreducible representations with the same
$J$ but different $L$ is also ruled out, as the $\bmS^2$ and the spin-orbit
potentials cannot change the angular  momentum $L$. On the other hand, the
tensor potential can generate transitions among the $\Gamma^{S=1}_{L\pm 1}$
fully symmetric currents by two units of $L$.

\subsection*{Anomalous Dimensions}

\begin{figure}[t] 
\begin{center}
 \leavevmode
 \epsfxsize=5cm
 \leavevmode
 \epsffile{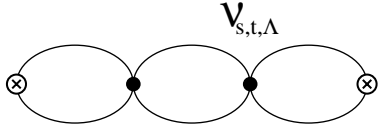}
 \vskip  0.0cm
 \caption{
Three-loop current correlator diagrams with insertions of the spin-dependent
potentials for the computation of the NLL anomalous dimension of the
currents. Combinatorial factors are suppressed.
 \label{figspincorrelator} }
\end{center}
\end{figure}

The spin-orbit and tensor potentials defined in Eq.~(\ref{tensordef1})
contribute to the NLL anomalous dimension of the spin triplet currents through
insertions in the three-loop current correlator diagrams in
Fig.\,\ref{figspincorrelator}. A single insertion of the spin-orbit potential
$-i{\cal V}_{\Lambda}^{(s)} S^{[ij]} \cdot T^{[p^i {p'}^j]}/(2m^2\bmk^2)$ 
yields for any of the spin triplet currents defined in
Eqs.~(\ref{tripletcurrent1},\ref{tripletcurrent2},\ref{tripletcurrent3})

%
%
\begin{eqnarray}
\lefteqn{
-i\,m\,
N_c \, {\cal V}_c^{(s)}\,{\cal V}_{\Lambda}^{(s)}
\int D^n \bmq_1\, D^n \bmq_2\, D^n \bmq_3 \,
}\nn\\ && \times
 \frac{\mbox{Tr}\Big[
\left( 
\Gamma_{J}^{S=1}(\bmq_3,\bmsigma)^{i_1\dots i_{J}}(i\sigma_2)
\right)^\dagger\,[S^i,S^j]\,
\left(
 \Gamma_{J}^{S=1}(\bmq_1,\bmsigma)^{i_1\dots i_{J}}(i\sigma_2)
\right)
\Big]q_2^i\,q_3^j}
{(\bmq_1^2+\delta)\,(\bmq_1-\bmq_2)^2\,(\bmq_2^2+\delta)\,
(\bmq_2-\bmq_3)^2\,(\bmq_3^2+\delta)}\,,    
\label{Vspinorbitcorrelator}
\end{eqnarray}
where the proper contraction of spinor indices is in analogy to
Eqs.\,(\ref{S2op},\ref{SLop}). Note that the correlator of two spin triplet 
currents with the same $J$ but different $L$ vanishes, as the spin-orbit
potential commutes with the squared angular momentum operator $\bmL^2$.  To
compute the contraction in the numerator of Eq.~(\ref{Vspinorbitcorrelator})
it is useful to perform the shift (\ref{shift}) in the $\bmq_1$
integration. This reduces all non-trivial contractions in the numerator to the
form of Eq.~(\ref{TyTx}) after use of the relations (\ref{Trelations}). The
computation then reduces to  
\begin{eqnarray}
&&-i m\,N_c\,{\cal V}_c^{(s)}\,
  {\cal V}_{\Lambda}^{(s)}\frac{N_L\,c_{J,L}}{2L+n-2}\,\mbox{Tr}\,[\bmone]\nn\\ 
&&\times
\int D^n \bmq_1\, D^n \bmq_2\, D^n \bmq_3 \, 
\frac{|\bmq_1|^L|\bmq_2||\bmq_3|^{L+1} P_L(n,\bar\bmq_1.\bar\bmq_2)
\Big(P_{L+1}(n,\bar\bmq_2.\bar\bmq_3)-P_{L-1}(n,\bar\bmq_2.\bar\bmq_3)\Big)}
{(\bmq_1^2+\delta)\,(\bmq_1-\bmq_2)^2\,(\bmq_2^2+\delta)\,
(\bmq_2-\bmq_3)^2\,(\bmq_3^2+\delta)}\,,\qquad
\label{Vspinorbitcorrelator2}
\end{eqnarray}
with
\begin{eqnarray}
c_{J,L}=\left\{ \begin{array}{rl} -L(L+1)(2L+n)\frac{(L+n-2)}{(2L+n-2)}\,,  & 
                 \;\;\; J=L+1 \\
         L(n-2)(L+n-2)\,,  & \;\;\; J=L \\
	 L+n-2\,, & \;\;\; J=L-1
	 \end{array} \right.
\end{eqnarray}

The calculation of the three-loop current correlator with a tensor potential
insertion $-i{\cal V}_t^{(s)}(-2n)S^{ij}T^{ij}(\bmk)/(m^2\bmk^2)$ can be
carried out similarly. For the case where the triplet currents at the sides
have the same $L$ the correlator reads 
\begin{eqnarray}
\lefteqn{
-im\,N_c\, {\cal V}_c^{(s)}\,{\cal V}_{t}^{(s)}(-2n)
\int D^n \bmq_1\, D^n \bmq_2\, D^n \bmq_3 \, 
}
\nn\\ && \times
\frac{\mbox{Tr}\Big[
\left(
 \Gamma_{J}^{S=1}(\bmq_3,\bmsigma)^{i_1\dots i_{J}}(i\sigma_2)
\right)^\dagger\,S^{ij}\,T^{ij}(\bmq_3-\bmq_2)\,
\left(
 \Gamma_{J}^{S=1}(\bmq_1,\bmsigma)^{i_1\dots i_{J}}(i\sigma_2)\right)
\Big]}
{(\bmq_1^2+\delta)\,(\bmq_1-\bmq_2)^2\,(\bmq_2^2+\delta)\,(\bmq_2-\bmq_3)^2\,
(\bmq_3^2+\delta)}\nn\\[3mm]
& =& -i m\,N_c\,{\cal V}_c^{(s)}\,{\cal V}_{t}^{(s)}N_L\,\tilde{c}_{J,L}\,
  \mbox{Tr}\,[\bmone]
 \int D^n \bmq_1\, D^n \bmq_2\, D^n \bmq_3 \, 
 \frac{|\bmq_1|^L|\bmq_3|^{L} P_L(n,\bar\bmq_1.\bar\bmq_2)\,
  \chi_L(n;\bmq_2,\bmq_3)}
{(\bmq_1^2+\delta)\,(\bmq_1-\bmq_2)^2\,(\bmq_2^2+\delta)\,(\bmq_3^2+\delta)}
\,,\nn\\&&
\label{Vtensorcorrelator}
\end{eqnarray}
with
\begin{eqnarray}
\chi_L(n;\bmp,\bmq)&=& (n-2)(L+n-2)\,P_{L}(n,\bar\bmp.\bar\bmq) \nn\\[3mm]
&&-\,n\,\frac{(n-2)(L+n-2)}{2L+n-2}
\frac{|\bmp||\bmq|}{(\bmp-\bmq)^2}
\Big(P_{L-1}(n,\bar\bmp.\bar\bmq)-P_{L+1}(n,\bar\bmp.\bar\bmq)\Big) \,,
\end{eqnarray}
and
\begin{eqnarray}
\tilde{c}_{J,L}=\left\{ \begin{array}{cl} \frac{2L(L+1)}{2L+n-2}\,,  & \;\;\; J=L+1 \\
         -2L\,,  & \;\;\; J=L \\
	 \frac{2}{2L+n-4}\,, & \;\;\; J=L-1
	 \end{array} \right.
\end{eqnarray}
The divergent part of Eqs.~(\ref{Vspinorbitcorrelator2})
and~(\ref{Vtensorcorrelator}) can be extracted from the integrals listed in
Appendix D. The insertion of a $\bmS^2$ potential is identical to that of the
${\cal V}_2/m^2$ potential, Eq.~(\ref{V2correlatorexample}), and the result
can be read off from  Eq.~(\ref{Zscalarpot}). The result for the
renormalization constant of the currents from the diagrams in
Fig.~\ref{figspincorrelator} including combinatorial factors thus reads  
($S\equiv 1$)
\begin{eqnarray}
\delta z^{\rm NLL}_{c,spin} 
& = & 
-\frac{{\cal V}_c^{(s)}(\nu)\,{\cal V}_{\Lambda}^{(s)}(\nu)}{256\pi^2}
\,\frac{J(J+1)-L(L+1)-2}{(2L+1)L(L+1)}\,(1-\delta_{L0})\nn\\
&&
-\frac{3{\cal V}_{c}^{(s)}(\nu)\,{\cal V}_t^{(s)}(\nu)}{64\pi^2}\,
 \frac{f_{J,L}}{2L+1}\,(1-\delta_{L0})
-\, \frac{{\cal V}_c^{(s)}(\nu){\cal
    V}_s^{(s)}(\nu)}{64\pi^2}\,\delta_{L0}\,S(S+1)\,,
\label{Zspin0}
\end{eqnarray}
\begin{eqnarray}
f_{J,L}=\left\{ \begin{array}{cl} -\frac{1}{J(2J+1)}\,,  & \;\;\; J=L+1 \\
         \frac{1}{J(J+1)}\,, & \;\;\; J=L \\
	 -\frac{1}{(J+1)(2J+1)}\,, & \;\;\; J=L-1
	 \end{array} \right.
\label{Zspin}	 
\end{eqnarray}
Explicit expressions for the Wilson coefficients of the spin-dependent
potentials can be found in Ref.~\cite{amis}. 
For ${}^{2S+1}L_J={}^3S_1$ the result agrees with the known results 
from Refs.~\cite{LMR,Hoang3loop,amis3,Pineda2}, and for 
${}^{2S+1}L_J={}^3P_{1,0}$ the renormalization constant at the
matching scale $\nu=1$ is consistent with the IR-divergences in the hard
region obtained from the threshold expansion in
Ref.~\cite{2loop_hard}. 

\begin{figure}[t] 
\begin{center}
 \leavevmode
 \epsfxsize=6cm
 \leavevmode
 \epsffile{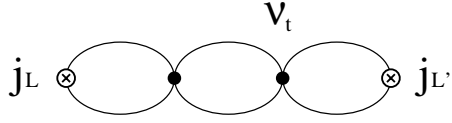}
 \vskip  0.0cm
 \caption{
Three-loop current correlator diagram between the currents with
orbital angular momentum $L$ and $L'=L\pm 2$ and a tensor potential insertion.
 \label{figmixcorrelator} }
\end{center}
\end{figure}

The tensor potential can also change the orbital angular momentum $L$ by two
units while conserving the total angular momentum $J$.\,\footnote{Note, however,
  that the spin of the fermion-antifermion system (S=0,1) cannot be changed by
  the $S=2$ tensor $S^{ij}$. Changes of the angular momentum $L$ by one unit
  are not allowed due to parity.}
Thus transitions between spin triplet currents with orbital angular momenta
$L=J\mp 1$ and $L'=J\pm 1 =L\pm 2$  
driven by diagrams such as Fig.~\ref{figmixcorrelator} with a tensor potential
insertion  do not vanish and could induce a mixing between the currents
through renormalization. The expression for the three-loop correlator  
of Fig.~\ref{figmixcorrelator}, where the tensor potential is inserted next to
the annihilation current on the RHS, reads now
%
%
\begin{eqnarray}
\lefteqn{
-i m\,N_c\, {\cal V}_c^{(s)}\,{\cal V}_{t}^{(s)}(-2n)
\int D^n \bmq_1\, D^n \bmq_2\, D^n \bmq_3 \, 
}
\nn\\ && \times
\frac{\mbox{Tr}\Big[
\left(
 \Gamma_{J,L'}^{S=1}(\bmq_3,\bmsigma)^{i_1\dots i_{J}}(i\sigma_2)
\right)^\dagger\,S^{ij}\,T^{ij}(\bmq_3-\bmq_2)\,
\left(
 \Gamma_{J,L}^{S=1}(\bmq_1,\bmsigma)^{i_1\dots i_{J}}(i\sigma_2)
\right)
\Big]}
{(\bmq_1^2+\delta)\,(\bmq_1-\bmq_2)^2\,(\bmq_2^2+\delta)\,
(\bmq_2-\bmq_3)^2\,(\bmq_3^2+\delta)}\nn\\[3mm]
& = &
-i m\,N_c\,{\cal V}_c^{(s)}\,{\cal
  V}_{t}^{(s)}(2n\,J\,N_{J+1})\,\mbox{Tr}\,[\bmone]
\nn\\ & & \times
 \int D^n \bmq_1\, D^n \bmq_2\, D^n \bmq_3 \, 
 \frac{|\bmq_1|^L|\bmq_3|^{L'} P_L(n,\bar\bmq_1.\bar\bmq_2)\,
\tilde{\chi}_{J,L}(n;\bmq_2,\bmq_3)}
{(\bmq_1^2+\delta)\,(\bmq_1-\bmq_2)^2\,(\bmq_2^2+\delta)\,
(\bmq_2-\bmq_3)^2\,(\bmq_3^2+\delta)}
\label{Vtensorcorrelatormix}
\end{eqnarray}
with
\begin{eqnarray}
\tilde\chi_{J,L}(n;\bmp,\bmq) &=& 
 \left\{ \begin{array}{cl}
  \left( \bmp^2 \,P_{J+1}(n,\bar\bmp.\bar\bmq)-2 |\bmp||\bmq| P_{J}(n,\bar\bmp.\bar\bmq)
        + \bmq^2 \,P_{J-1}(n,\bar\bmp.\bar\bmq) \right)\,,  & \;\;\; L=J-1 \nn\\[3mm]
  \left( \bmp^2 \,P_{J-1}(n,\bar\bmp.\bar\bmq)-2 |\bmp||\bmq| P_{J}(n,\bar\bmp.\bar\bmq)
        + \bmq^2 \,P_{J+1}(n,\bar\bmp.\bar\bmq) \right)\,, & \;\;\; L=J+1 
	 \end{array} \right. \,.\nn\\
\end{eqnarray}
The divergent contributions of Eq.~(\ref{Vtensorcorrelatormix}) can be
obtained from the results for the relevant integrals given in Appendix D. 
We find that the UV divergences from the sum of both insertions of the tensor
potential cancel and that there is no mixing between the $j_{J=L-1}^{S=1}$ and
the $j_{J=L+1}^{S=1}$ currents at NLL order.

\section{Solutions and Discussions} 
\label{sectiondiscussion}

\subsection*{Scalar Production Currents}

For scalar fields, only the spin-independent contributions shown in
Eq.~(\ref{Zscalarpot}) have to be taken into account and we can write the RGE
for the Wilson coefficient of the current with quantum number $L$ in terms  
of the RGE's of the S- and P-wave current Wilson coefficients which
were determined in Ref.~\cite{HoangRuizSvNRQCD}, 
\begin{eqnarray}
\frac{\partial}{\partial\ln \nu}\ln[c_L(\nu)] & = & 
  4\delta z^{\rm NLL}_c \nn\\ 
&=& \delta_{L0}\,\frac{\partial}{\partial\ln \nu}\ln[c_S(\nu)]+
\frac{3}{2L+1}\left(1-\delta_{L0}\right)\frac{\partial}{\partial\ln
  \nu}\ln[c_P(\nu)]\,. 
\label{scalarRGE1}
\end{eqnarray}
The solution is thus found
immediately and reads 
\begin{eqnarray}
\ln \frac{c_L(\nu)}{c_L(1)}&=& B_2\,\pi \alpha_s(m)\, (1-z) + B_3\,\pi \alpha_s(m)\, \ln (z) 
 \nn\\[3pt] &&   + \,B_4\,\pi \alpha_s(m)\,\Big[1-z^{1-13C_A/(6\beta_0)} \Big] 
  + B_0\,\pi \alpha_s(m)\Big[z-1-w^{-1}\ln (w)\Big]
\,,
\label{runningscalar}
\end{eqnarray}
where $z=\alpha_s(m\nu)/\alpha_s(m)$,
$w=\alpha_s(m\nu^2)/\alpha_s(m\nu)$, and
\begin{eqnarray}
B_i=\delta_{L0}\,b_i+\frac{3}{2L+1}\left(1-\delta_{L0}\right)\,d_i\,,
\end{eqnarray}
and with~\cite{HoangRuizSvNRQCD}
\begin{eqnarray}
b_2 & = & \frac{C_F \,[\, C_A C_F(9C_A-100C_F)-\beta_0(26C_A^2+19C_F
    C_A -32C_F^2)\,]\,}{26\beta_0^2 C_A} 
\,,\nn\\[3mm]
b_3 & = & 
-\frac{C_F^2\,[\,C_A(9C_A - 100 C_F) - 6 \beta_0(3C_A - 4C_F)\,]}{2 \beta_0^2(6\beta_0 - 13 C_A)}
\,,\nn\\[3mm] 
b_4 & = & \frac{24C_F^2(11C_A-3\beta_0)(5C_A+8C_F)}{13C_A(6\beta_0-13C_A)^2} 
\,,\nn\\[3mm] 
b_0 & = & - \frac{8C_F(C_A+C_F)(C_A+2C_F)}{3\beta_0^2}
\,,\nn\\[3mm]
d_2 & = & -\frac{ C_F (C_A+2C_F)}{3\beta_0}
\,,\nn\\[3mm]
d_0 & = & -\frac{8 C_A C_F(C_A+4C_F)}{9\beta_0^2}\,.
\end{eqnarray}

\begin{figure}
\epsfxsize=10.cm 
\epsffile[80 450 560 720]{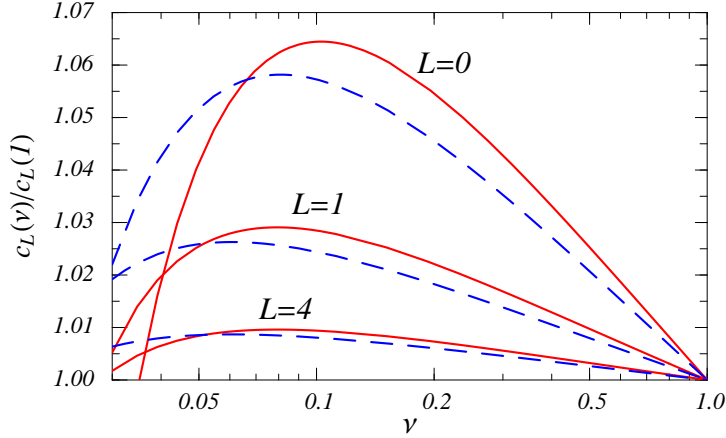}
{\caption{NLL running of the normalized Wilson coefficient 
    for colored scalars for
    $L=0,1$ and $4$ for $m=220$~GeV (solid lines) and
    for $m=500$~GeV (dashed lines).}
\label{singletplot}}
\end{figure}

In Fig.~{\ref{singletplot}} we have displayed the NLL running of the 
scalar currents exemplarily for $L=0,1$ and $4$ normalized to their
matching values. For the input parameters we have chosen  
two different values for the heavy scalar mass, $m=220$~GeV (solid lines) 
$m=500$~GeV (dashed lines), and $\alpha_s(m_Z)=0.118$, taking 
leading-logarithmic running for $\alpha_s$ with $n_f=5$ active massless 
quark flavors and no active massless squarks ($n_s=0$). As is already 
visible from the form of Eq.~(\ref{scalarRGE1}), we find that the 
$\nu$-variation of the currents becomes weaker for larger angular 
momenta. While the S-wave coefficient increases by about 6\% for 
$\nu\sim\alpha_s$ compared to the matching value at $\nu=1$, the
P-wave coefficient only increases by less than 3\%. For $L\le 4$
the maximal relative variation is already below 1\%.
It is also conspicuous that for any $L$ the maximum slightly
decreases with the heavy scalar mass and also moves towards smaller
values of $\nu$. As already discussed in Ref.~\cite{HoangRuizSvNRQCD}, the
latter feature can be understood qualitatively from the fact that the average
velocity $\langle  v\rangle\sim\alpha_s(m\alpha_s)$ decreases with the heavy
scalar mass $m$.  

\subsection*{Fermion Production Currents}

The RGE of Eq.~(\ref{rgeCurrent}) for the Wilson coefficients of the currents which
produce a pair of quarks with quantum numbers $J,\,L$ and $S$ is obtained at
NLL by adding the anomalous dimensions from the diagrams with spin independent
and spin dependent potentials (Eqs.~(\ref{Zscalarpot}) and (\ref{Zspin0})
respectively):   
\begin{eqnarray}
\frac{\partial}{\partial\ln\nu}\ln[c_{J,L}^S(\nu)]& = & 
4\delta z^{\rm NLL}_c+ 4\delta z^{\rm NLL}_{c,spin}\nn\\ 
& = &
\delta_{L0}\,\frac{\partial}{\partial\ln\nu}\ln[c_{J,0}^S(\nu)]\nn\\
&&+\,\frac{(1-\delta_{L0})}{2L+1}\bigg\{
-\frac{1}{16\pi^2}\,{\cal V}_c^{(s)}(\nu)\,
\Big( \frac{{\cal V}_c^{(s)}(\nu)}{4} + {\cal V}_r^{(s)}(\nu) \Big)
\nonumber\\[2mm] & & \qquad +\,
\alpha_s(m\nu)^2 \Big(3{\cal V}_{k1}^{(s)}(\nu)+2{\cal V}_{k2}^{(s)}(\nu) \Big)
+\,
\frac{\alpha_s^2(m\nu)}{2}\Big(\frac{C_F^2}{2}-C_AC_F\Big)
\nonumber\\[2mm] & & \qquad  
 -\,\frac{{\cal V}_c^{(s)}(\nu)\,{\cal V}_{\Lambda}^{(s)}(\nu)}{64\pi^2}
\,\frac{J(J+1)-L(L+1)-2}{L(L+1)}\,\delta_{S1}
\nonumber\\[2mm] & & \qquad 
-\frac{3{\cal V}_{c}^{(s)}(\nu)\,{\cal V}_t^{(s)}(\nu)}{16\pi^2}\,
 f_{J,L}\,\delta_{S1}
\,\bigg\}
\,.
\label{rgeCJL}
\end{eqnarray}
We have singled out the known RGE for the S-wave
currents~\cite{LMR,HoangStewartultra,Pineda2}, 
$c_{J,0}^S$, that is relevant for $\gamma^* \to q\bar{q}$ production in the
spin triplet case, and for  $\gamma\gamma \to q\bar{q}$ in the spin singlet
case. Solving Eq.~(\ref{rgeCJL}) yields 
($z=\alpha_s(m\nu)/\alpha_s(m), w=\alpha_s(m\nu^2)/\alpha_s(m\nu)$)
\begin{eqnarray}
\ln \frac{c_{J,L}^S(\nu)}{c_{J,L}^S(1)}&=& A_2\,\pi \alpha_s(m)\, (1-z) + 
A_3\,\pi \alpha_s(m)\, \ln (z) 
 \nn\\[2pt] &&   + \,A_4\,\pi \alpha_s(m)\,\Big[1-z^{1-13C_A/(6\beta_0)} \Big] 
  + \,A_5\,\pi \alpha_s(m)\,\Big[1-z^{1-2C_A/\beta_0} \Big] 
 \nn\\[3pt] &&   + \,A_8\,\pi \alpha_s(m)\,\Big[1-z^{1-C_A/\beta_0} \Big] 
  + A_0\,\pi \alpha_s(m)\Big[z-1-w^{-1}\ln (w)\Big]
\,,
\label{runningfermion}
\end{eqnarray}
with
\begin{eqnarray}
A_2 &=& a_2\,\delta_{L0} - \frac{(1-\delta_{L0})}{2L+1}\,
\bigg\{\frac{C_F(C_A+2C_F)}{\beta_0}
+\frac{C_F^2}{4\beta_0}\frac{J(J+1)-L(L+1)-2}{L(L+1)}\,\delta_{S1}\,\bigg\}
\,,\nn\\[3mm]
A_3 &=& a_3\,\delta_{L0}\,,\nn\\[3.5mm]
A_4 &=& a_4\,\delta_{L0}\,,\nn\\[3mm]
A_5 &=& a_5\,\delta_{L0} +
\frac{(1-\delta_{L0})}{2L+1}\frac{f_{J,L}\,C_F^2}{2(\beta_0-2C_A)}\,\delta_{S1}
\,,\nn\\ 
A_8 &=&
\frac{(1-\delta_{L0})}{2L+1}\,\frac{C_F^2}{\beta_0-C_A}\,
\frac{J(J+1)-L(L+1)-2}{L(L+1)}\,\delta_{S1}
\,,\nn\\  
A_0 &=& a_0\,\delta_{L0} - \frac{(1-\delta_{L0})}{2L+1}\,
\frac{8C_A C_F(C_A+4C_F)}{3\beta_0^2} \,.
\end{eqnarray}
The coefficients $a_i$ for the running of the S-wave Wilson coefficients
$c_{J,0}^{S=0,1}$ read~\cite{HoangStewartultra,Pineda2} 
\begin{eqnarray} 
  a_2 &=& b_2\,,\quad 
  a_4 =  b_4\,,\quad 
  a_0  =  b_0
  \,, \nn\\[3mm]
  a_3 &=&
    \frac{C_F^2 }{ \beta_0^2\, (6\beta_0-13C_A) (\beta_0-2C_A)}
    \, \Big\{ C_A^2 ( 9 C_A- 100 C_F )
    + \beta_0\,C_A \Big[ 74 C_F + C_A ( 13 {\bf S^2} - 42 ) \Big]\nn \\[3mm]
   &&\qquad - 6 \beta_0^2  \Big[ 2 C_F + C_A ( {\bf S^2} - 3 ) \Big]\,\Big\} 
   \,,\nn\\[3mm]
 a_5 &=& \frac{C_F^2 \big[ C_A( 15\! -\! 14\,{\bf S^2} ) 
    \!+\! \beta_0 (4{\bf S^2}\!-\!3) \big] }{ 6 (\beta_0\!-\!2 C_A)^2 }\,,
\label{acoeffs2}  
\end{eqnarray}
where ${\bf S}^2\equiv 2$ in the triplet case and 
${\bf S}^2\equiv 0$ for the singlet case.

\begin{figure}
\epsfxsize=8.1cm 
\epsffile[80 450 560 720]{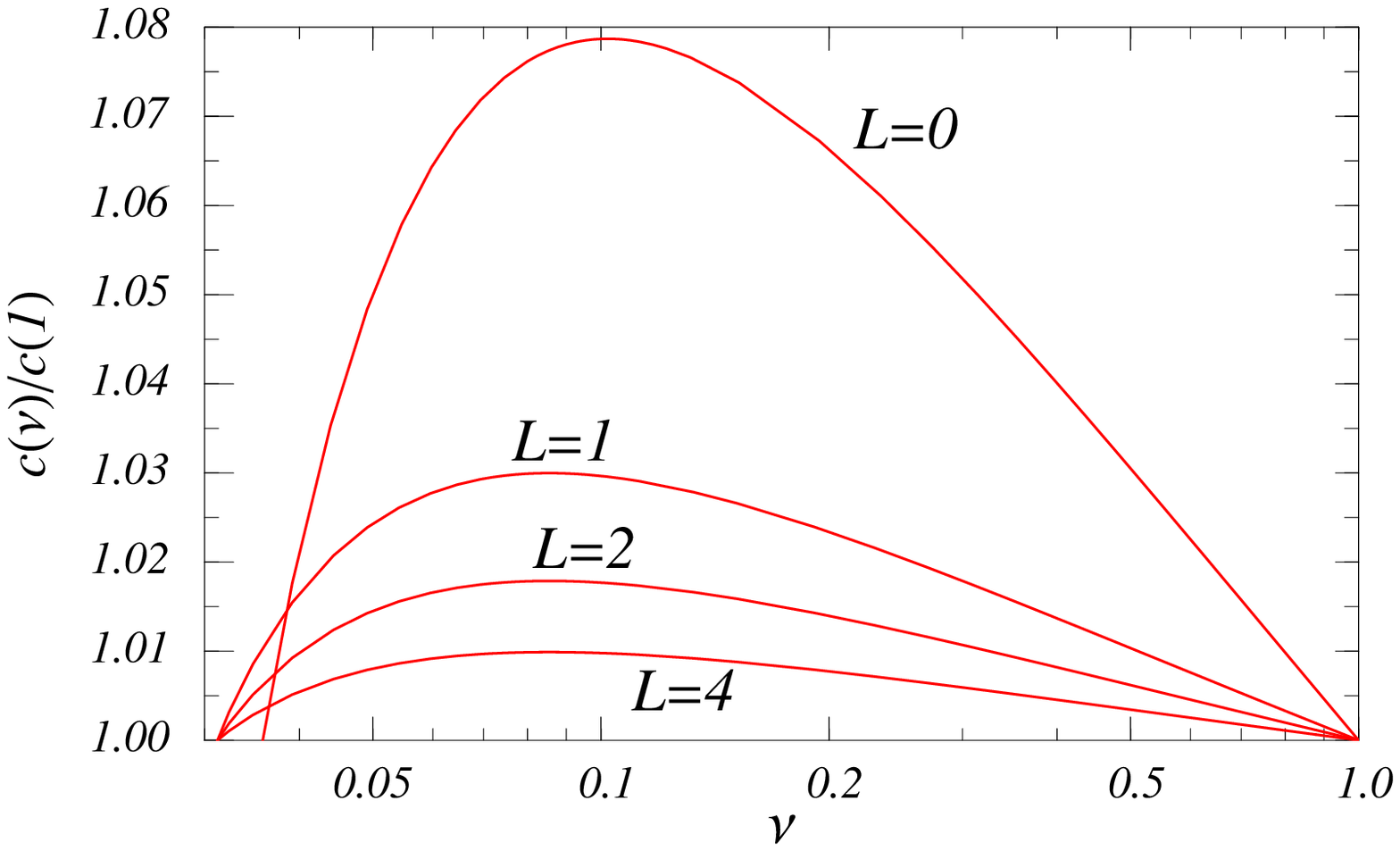}
\epsfxsize=8.1cm 
\epsffile[80 454 560 720]{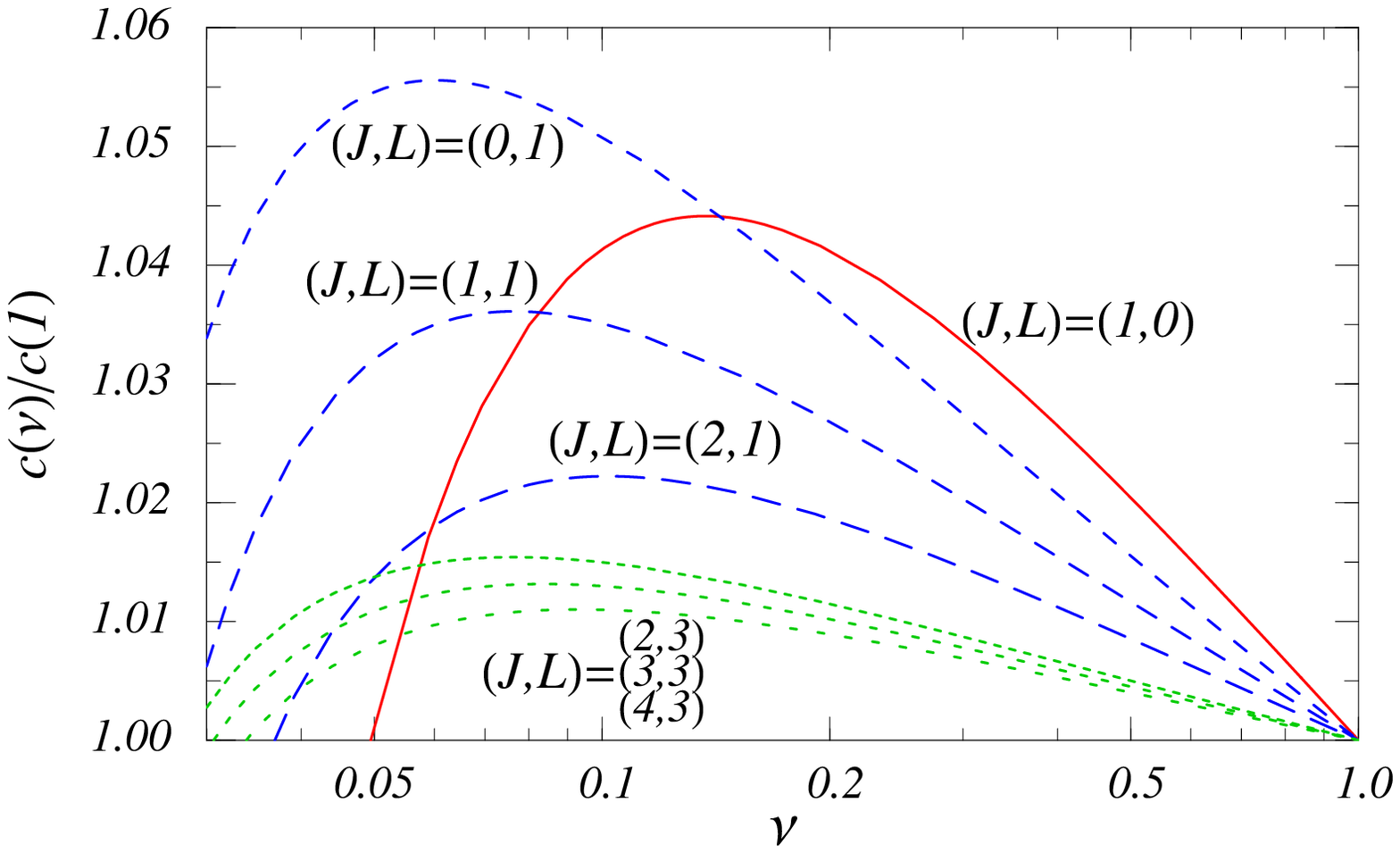}
{\caption{Left panel: NLL running of the normalized Wilson coefficient 
    for quarks in a spin singlet configuration for
    $L=0,1,2$ and $4$ for $m=175$~GeV. Right panel: NLL running for quarks in
    spin triplet configuration for $L=0,1,3$ and all possible $J$ values for
    $m=175$~GeV.}
\label{tripletplot}}
\end{figure}

In Fig.~{\ref{tripletplot}}a and b we have displayed the NLL running of the 
quark currents for spin singlets and triplets, respectively, for a number of
cases. For the heavy quark mass we have chosen $m=175$~GeV, and for the strong
coupling we employed the convention used in the discussion of the scalar
currents. For the singlet cases we have shown the evolution for
$L=0,1,2,4$. The dependence of the evolution of the currents on the value of
$L$ is very similar to the scalar currents.  
For the triplet cases we have displayed the evolution for the cases $L=0,1,3$ 
and all possible values of $J$. Here the dependence of evolution on the choice
of the angular momentum $L$ (for a fixed $J$ value) is similar to the spin
singlet cases. But we also find that the evolution for a fixed value of $L$ is
stronger for the smaller $J$ values.

\section{Comments on Scheme Dependent Results} 
\label{sectioncomments}

In this section we comment on scheme dependences of a number of
results for nonrelativistic currents that can be found in the
literature. The current that has been studied most extensively is the
vector current $\psi^\dagger_{\bmp}\bmsigma(i\sigma_2)\chi_{-\bmp}^*$, which
governs heavy quark pair production in $e^+e^-$
annihilation~\cite{FadinKhoze,Jezabek,Hoangsynopsis} or the large
Higgs energy endpoint 
region in $e^+e^-\to t\bar t H$~\cite{Farrell1,Farrell2}. At NNLL
order the matching condition and the anomalous dimension depend on
the scheme used for the spin-dependent interactions and
potentials. Using dimensional regularization such computations were
carried out in
Refs.~\cite{Czarnecki1,Beneke4,hmst1,Hoang3loop,HoangStewartultra}.   

In Refs.~\cite{hmst1,Hoang3loop,HoangStewartultra}, the NNLL order
matching condition of the vector current 
$\psi^\dagger_{\bmp}\bmsigma(i\sigma_2)\chi_{-\bmp}^*$ was determined using
three-dimensional $\sigma$ structures for the spin-dependent
potentials, as shown in Eq.~(\ref{spinpotdef1}), and three-dimensional 
$\sigma$ algebra. Although the use of the three-dimensional $\sigma$
algebra is a priori inconsistent, the results for the NNLL order
matching conditions derived in
Refs.~\cite{hmst1,Hoang3loop,HoangStewartultra} nevertheless represent 
a viable scheme. This scheme is related to the spin-dependent potentials 
in Eqs.~(\ref{ndimpot}) or (\ref{tensordef1}), with strict $n$-dimensional
$\sigma$ algebra, through multiplicative $n$-dependent factors. For the
vector current $\psi^\dagger_{\bmp}\bmsigma(i\sigma_2)\chi_{-\bmp}^*$ this
scheme has, for the tree level potentials, the form 
\begin{eqnarray}
 V^{\prime\prime}({\bmp},{\bmp^\prime}) & \simeq & 
g_s^2 (T^A \otimes \bar T^A) \, \bigg\{
 \frac{1}{\bmk^2}
 + \frac{{\bmp^2 + \bmp^{\prime 2}}}{2 m^2 \bmk^2} 
 + \left(1-\frac{5}{12}\,\epsilon\right)
 \,\left[
    \frac{n-3}{4m^2 } 
   +\frac{\bmS^2}{m^2 } \left(\frac{\bmS^2}{n}-1 \right)
 \right]
\bigg\}
\nn\\[2mm] &\simeq & 
g_s^2 (T^A \otimes \bar T^A) \, \bigg\{
 \frac{1}{\bmk^2}
 + \frac{{\bmp^2 + \bmp^{\prime 2}}}{2 m^2 \bmk^2} 
 + \left(1+\epsilon\right)
 \,\left[
   -\frac{\bmS^2}{3\,m^2}
 \right]
 \bigg\}
\,,
\label{3dimpot}
\end{eqnarray}
and can be derived from the entries in Tab.~\ref{tab2}.
We have dropped the spin-orbit and tensor potentials that were
displayed in Eqs.~(\ref{ndimpot}) or (\ref{tensordef1}) because they
do not contribute for the vector current
$\psi^\dagger_{\bmp}\bmsigma(i\sigma_2)\chi_{-\bmp}^*$. The first relation in
(\ref{3dimpot}) allows to compute the difference between the NNLL
matching condition obtained in Ref.~\cite{Hoang3loop} and the scheme where the
NNLL matching condition is identified with the contributions from the hard
region in the threshold expansion~\cite{Czarnecki1,Beneke4}, which implies the
use of the scheme for the potentials shown in Eq.~(\ref{ndimpot}) (see also
Ref.~\cite{Pineda:2006ri}). These two schemes only differ in the
treatment of the spin-dependent potentials. 
Using the contribution to the NLL anomalous
dimension from the spin-dependent potential ${\cal V}_s^{(s)}$ at the
hard matching scale from Eq.~(\ref{Zspin0}), 
$\delta z_{c,{\rm spin}}^{\rm NLL}=
-{\cal V}_c^{(s)}(1){\cal
  V}_s^{(s)}(1)/32\pi^2=C_F^2\alpha_s^2(m)/6$~\cite{amis,amis2},
one finds that the relation between the NNLL matching condition based
on the three-dimensional $\sigma$ algebra from Ref.~\cite{Hoang3loop},
$c_{\rm NNLL,v}^{\rm 3\,dim}(1)$, and the one based on the threshold
expansion scheme, $c_{\rm NNLL,v}^{\rm thr\,exp}(1)$, reads
\begin{eqnarray}
c_{\rm NNLL,v}^{\rm 3\,dim}(1) & = &
 c_{\rm NNLL,v}^{\rm thr\,exp}(1) - \frac{5}{12}\delta z_{c,{\rm
 spin}}^{\rm NLL}
\, = \,
 c_{\rm NNLL,v}^{\rm thr\,exp}(1) - \frac{5}{72}\,C_F^2\alpha_s^2(m)
\,,
\label{crelations}
\end{eqnarray}
which agrees with the results given in
Refs.~\cite{Czarnecki1,Beneke4,Hoang3loop}. 
It is also straightforward to compute such a scheme relation
for the spin-dependent non-mixing contributions of the
NNLL anomalous dimension of the vector current that was also computed 
in Ref.~\cite{Hoang3loop}.

Scheme relations for the NNLL matching conditions analogous to
Eq.~(\ref{crelations}), connecting the scheme from Ref.~\cite{Hoang3loop} 
with the one based on the threshold expansion can also be derived for the 
${}^3P_1$ axial-vector 
($J^i_{av}=\psi^\dagger_{\bmp}\frac{p^\ell}{m}\sigma^{\ell i}(i\sigma_2)\chi_{-\bmp}^*$),
${}^3P_0$ scalar
($J_{s}=\psi^\dagger_{\bmp}\frac{p^\ell}{m}\sigma^{\ell}(i\sigma_2)\chi_{-\bmp}^*$),
and ${}^1S_0$  pseudo-scalar 
($J_{ps}=\psi^\dagger_{\bmp}(i\sigma_2)\chi_{-\bmp}^*$)
currents:
\begin{eqnarray}
c_{\rm NNLL,av}^{\rm 3\,dim}(1) & = &
 c_{\rm NNLL,av}^{\rm thr\,exp}(1) - \frac{1}{16}\,C_F^2\alpha_s^2(m)
\,,
\\[2mm]
c_{\rm NNLL,s}^{\rm 3\,dim}(1) & = &
 c_{\rm NNLL,s}^{\rm thr\,exp}(1) - \frac{1}{24}\,C_F^2\alpha_s^2(m)
\,,
\\[2mm]
c_{\rm NNLL,ps}^{\rm 3\,dim}(1) & = &
 c_{\rm NNLL,ps}^{\rm thr\,exp}(1)
\,.
\end{eqnarray}
The respective NNLL matching conditions in the scheme based on the threshold
expansion were computed in Ref.~\cite{2loop_hard}.

Finally, we would like to make a few comments on the conventions for
$\gamma_5$. In previous NRQCD literature based on dimensional
regularization, only the totally anticommuting
version with the form $\gamma_5=(^{0\,\bmI}_{\bmI\,0})$ has been
considered~\cite{hmst1,2loop_hard}, which leads to the nonrelativistic
currents $\psi^\dagger_{\bmp}(i\sigma_2)\chi_{\bmp}^*$ and
$\psi^\dagger_{\bmp}\frac{p^\ell}{m}\sigma^{\ell i}(i\sigma_2)\chi_{\bmp}^*$ 
for the pseudo-scalar $\bar\psi\gamma_5\psi$ and axial-vector
$\bar\psi\gamma^i\gamma_5\psi$ ($i=1,\dots,n$) full theory currents,
respectively.  In particular, in Ref.~\cite{2loop_hard}
the NNLL (two loop) matching conditions for these pseudo-scalar and
axial-vector currents were identified with the hard contributions
obtained in the threshold expansion using the form
$\gamma_5=i\gamma^0\gamma^1\gamma^2\gamma^3=
-i\gamma_0\gamma_1\gamma_2\gamma_3$ for triangle graphs and the 
totally anticommuting $\gamma_5$ for the other diagrams. Such an
approach is viable as long as the
hard contributions from triangle graphs only lead to IR-finite terms. 
However, from the point of view of the full theory it
can be important or even mandatory to employ only the fully consistent
definition $\gamma_5=i\gamma^0\gamma^1\gamma^2\gamma^3$. The latter 
form for $\gamma_5$ 
also leads to a different from of the EFT currents from the
nonrelativistic expansion. Writing $\gamma_5=
\frac{i}{3!}\epsilon^{ijk}\gamma^0\gamma^{i}\gamma^{j}\gamma^{k}=
\frac{i}{3!}\epsilon^{ijk}\gamma^0\gamma^{ijk}$, and
$\gamma^i\gamma_5=\frac{i}{2}\epsilon^{ijk}\gamma^0\gamma^{jk}$,
where $\epsilon^{ijk}$ is the usual three-dimensional
$\epsilon$-tensor if all indices assume values $1,2$ or $3$, and
identically zero otherwise, one obtains for the leading order
expansion 
\begin{eqnarray}
\bar\psi\gamma_5\psi 
& \to &
 -\frac{i}{3!}\,\epsilon^{ijk}\,
  \psi^\dagger_{\bmp}\sigma^{ijk}(i\sigma_2)\chi_{-\bmp}^*
\,,
\label{altcur1}
\\[2mm]
\bar\psi\gamma^i\gamma_5\psi 
& \to &
 -\frac{i}{2}\,\epsilon^{ijk}\,
  \psi^\dagger_{\bmp}\left[\frac{1}{m}(p^j\sigma^k-p^k\sigma^j)\right]
 (i\sigma_2)\chi_{-\bmp}^*
\,.
\label{altcur2}
\end{eqnarray} 
The current in Eq.~(\ref{altcur1}) was discussed in
Sec.~\ref{sectiontriplet} and the one in Eq.~(\ref{altcur2}) is a
simple example for the alternative ${}^3L_L$ currents involving the
tensor $\widetilde\Gamma^{S=1}_L$ shown in Eq.~(\ref{newLcurrent}).
The EFT currents on the RHS's of Eqs.~(\ref{altcur1}) and
(\ref{altcur2}) represent a viable scheme and are equivalent to the
currents above for $n=3$, but inequivalent otherwise. So at NNLL order their
respective matching conditions and anomalous dimensions differ. Using
$\gamma_5=i\gamma^0\gamma^1\gamma^2\gamma^3$ in the full theory, the 2-loop
hard contributions from the threshold expansion give the NNLL matching
conditions of the currents in Eqs.~(\ref{altcur1}) and (\ref{altcur2}).

\section{Conclusion}
\label{sectionconclusion}
The construction of non-relativistic interpolating currents for
describing the production and annihilation of a color singlet heavy
particle-antiparticle pair with arbitrary quantum numbers
${}^{2S+1}L_J$ in $n=3-2\epsilon$
dimensions has been addressed in this work. Arbitrary angular momentum
configurations are accounted for by the generalization of the
spherical harmonics in $n$ dimensions, for which we have provided
a representation in terms of cartesian coordinates. The use of the
$n$-dimensional spherical harmonics is required to maintain SO(n)
rotational invariance in calculations within dimensional
regularization.

Similarly, a consistent description of the spin of a
fermion-antifermion pair requires the treatment of Pauli
$\sigma$-matrices in $n$ dimensions. The antisymmetrized product of
$\sigma$-matrices are the proper irreducible tensors with respect to
SO(n) to build interpolating currents describing an arbitrary
angular-spin state of a fermion-antifermion pair. In $n$ dimensions
the form of irreducible currents in SO(n) which reduce to a specific
angular-spin state in $n=3$ is not unique. This is related to the
existence of evanescent operators that can be built from the
nonrelativistic field operators and the $\sigma$-matrices in $n$
dimensions. Such evanescent structures occur even in simple standard
processes, and in general need to be taken into account consistently
at subleading order for the renormalization process.
The specific choice of a basis for the physical currents defines a
specific renormalization scheme. Similar scheme choices are possible
for the  spin-dependent interactions and potentials. In this work we
have discussed different versions for spin-singlet and spin-triplet
currents, and also for the spin-dependent potential, which are
inequivalent in general $n$ dimensions, but equivalent for $n=3$.

We have discussed the latter issues in the framework of NRQCD and have
shown that at NNLL order transitions between currents containing
different irreducible $\sigma$ structures cannot occur. Since for a
specific angular-spin state the corresponding leading order currents
do not have an anomalous dimension at leading-logarithmic order, the
scheme-dependence introduced by specific choices for
the currents and spin-dependent potentials in NRQCD only affects NNLL
matching conditions and anomalous dimensions. The relation between NNLL
matching conditions of a number of currents for schemes used in the
literature have been determined.

Finally, we have determined the NLL order anomalous dimensions for
the leading interpolating currents for quarks-antiquark and colored
scalar-antiscalar pairs for arbitrary spin and angular momentum
$(J,L,S)$ quantum numbers. We have shown by explicit computation that 
the spin-dependent potential do not cause any mixing between currents  
with different $L$ at NLL order.
An important property of the solution of the respective anomalous
dimensions is that the scale variation of the currents becomes in
general weaker for higher angular momentum $L$.

\begin{acknowledgments} 
We thank A.~Buras, D.~Maison, M.~Misiak and U.~Nierste for discussions.
We particularly thank P.~Breitenlohner for numerous
helpful discussions.

\end{acknowledgments}

\mbox{}
\vskip 1cm

\appendix

\section{Fermion pair production from $\gamma\gamma$ collisions}
\label{appendix2}
\begin{figure}
\begin{center}
\includegraphics[width=4cm]{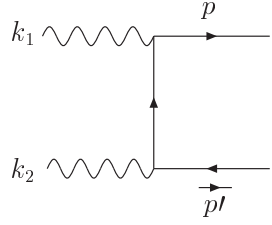}
\end{center}
\caption{Fermion pair production from two photons. \label{fig:2photon}}
\end{figure}

The $\gamma\gamma \to f\bar{f}$ amplitude at lowest order in $\alpha$ 
is given by the graph in Fig.~\ref{fig:2photon} plus the one 
with permutted photons.  
Let us write the production amplitude up to second order in 
the three-momentum of the fermions in the c.m.\,frame $\bmp$. The outgoing
momenta are $p=(E,\bmp)$, and $p^\prime=(E,-\bmp)$ with
$E=\sqrt{m^2+\bmp^2}=m+\frac{\bmp^2}{2m}+\dots$, and the three-momenta of the
photons are $\bmk_1=-\bmk_2\equiv\bmk$. 
We shall use Pauli spinors $\psi$ and $\chi$ for the fermion and the 
antifermion respectively,
\begin{eqnarray}
u(\bmp) = \sqrt{\frac{E+m}{2E}}
\left( \begin{array}{c} \psi_{\bmp} \\ \frac{\bm{\sigma}\cdot\bmp}{m+E}
\,\psi_{\bmp} \end{array} \right)\;\;\;,
\;\;\;
v(\bmp) = \sqrt{\frac{E+m}{2E}}\left( \begin{array}{c} 
\frac{\bm{\sigma}\cdot\bmp}{m+E}\,(i\sigma_2)\chi_{\bmp}^* \\
(i\sigma_2)\chi_{\bmp}^*  \end{array} \right)
\,,
\label{spinors}
\end{eqnarray}
and pick a gauge where
the photon polarizations $\epsilon_i$ are purely transverse,
$\epsilon_i\cdot k_i=0,\,\epsilon_i^0=0$. The result for the full
amplitude in Fig.~\ref{fig:2photon} including the crossed graph is 
\begin{eqnarray}
i{\cal M} = -i\,e^2Q_f^2\,\bar{u}(\bmp)\,\xslash{\epsilon_1}\, 
\frac{1}{\xslash{p}-\xslash{k}_1-m}\,
\xslash{\epsilon_2}\,\,v(-\bmp)+\{ 1\leftrightarrow 2\}\,,
\end{eqnarray}
and expanding in $\bmp$ gives
\begin{eqnarray}
i{\cal M}=i\frac{e^2Q_f^2}{m^2}\,\bar{u}(\bmp) \!\!\! &\Big[& \!\!\!
  \epsilon_1^i\epsilon_2^j 
\,\gamma^{ij\ell} 
 \left( -k^\ell + \frac{(\bmk\bmp)}{m^2}p^\ell \right) 
+\Big((\bm{\epsilon}_1\bmp)\epsilon_2^i + (\bm{\epsilon}_2\bmp)\epsilon_1^i
+\frac{(\bmk\bmp)}{m^2} (\bm{\epsilon}_1\cdot\bm{\epsilon}_2)k^i \Big)
\,\gamma^{i} \nn\\[3mm]
&&-\,m(\bm{\epsilon}_1\bm{\epsilon}_2)\,\left(m+\gamma^\ell p^\ell\right)
+\frac{(\bmk\bmp)}{m} \,\epsilon_1^i \epsilon_2^j\,\gamma^{ij}
\,\Big]v(-\bmp)\,+\,\ldots\,.
\label{B3}
\end{eqnarray}
The spinor structures in Eq.~(\ref{B3}) can be further expanded in $\bmp$ by using the 
leading order matching relations between full theory currents and the
non-relativistic ones shown in Table~\ref{tab1}:
\begin{eqnarray}
i{\cal M}&=&W_0^{ij\ell}\,\psi_{\bmp}^{\dagger}\,\sigma^{ij\ell}\,(i\sigma_2)\chi_{-\bmp}^*
+\frac{2}{n}\,W_1^{ii}\,\psi_{\bmp}^{\dagger}\,(\bmsigma\bmp)\,(i\sigma_2)\chi_{-\bmp}^*
\nn\\[3mm]
&&+ \, W_1^{ij}\,\psi_{\bmp}^{\dagger}\,\Big(p^i\, \sigma^j+p^j\,
\sigma^i-\frac{2}{n}\delta^{ij}(\bmsigma\bmp)\Big)\,
(i\sigma_2)\chi_{-\bmp}^*+{\cal O}(\bmp^2/m^2)\,,
\end{eqnarray}
where
\begin{eqnarray}
W_0^{ij\ell} &=& \frac{e^2Q_f^2}{m^2}\,\epsilon^i_1\epsilon^j_2 \,k^\ell \,,
\nn\\
W_1^{ij} &=& \frac{e^2Q_f^2}{2m^2}\, \Big[ \,  
(\bm{\epsilon}_1\bm{\epsilon}_2)\frac{k^i k^j}{m^2}
+ \epsilon_1^i\, \epsilon_2^j + \epsilon_1^j 
\epsilon_2^i \, \Big]\,.
\end{eqnarray}
Thus in fermion production from $\gamma\gamma$ collisions 
the currents 
$j_{L=0}^{[ij\ell]}$ (${}^1S_0$), 
$j^{S=1}_{L-1}$ (${}^3P_0$)
and
$(j^{S=1}_{L+1})^{ij}$ (${}^3P_2$)
are involved.


\section{Fermion-antifermion pair annihilation into $3\gamma$}
\label{appendix3}

\begin{figure}
\begin{center}
\includegraphics[width=5cm]{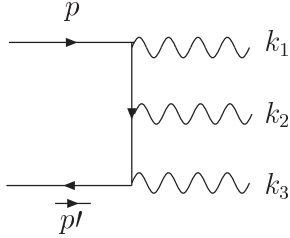}
\end{center}
\caption{Three photon annihilation graph. The $3!$ permutations of the photons
have to be considered. \label{fig:3photon}}
\end{figure}

The decay amplitude $f\bar{f}\to 3\gamma$ to lowest order in $\alpha$ 
is given by the graph in Fig.~\ref{fig:3photon} summed over the $3!$ permutations
of the photons. The amplitude of the graph shown is
\begin{eqnarray}
i{\cal M} = -ie^3Q_f^3\,\bar{v}(-\bmp)\,
\xslash{\epsilon_3}\,\, \frac{1}{-\xslash{p}^\prime+\xslash{k}_3-m}\,
\xslash{\epsilon_2}\,\, \frac{1}{\xslash{p}-\xslash{k}_1-m}\,
\xslash{\epsilon_1}\,\,u(\bmp)
\,.
\end{eqnarray}
We want to compute the leading term in the expansion of 
the three-momentum of the fermions in the c.m.\,frame $\bmp$. 
We can then work at threshold and the spinors reduce to
\begin{eqnarray}
u(\bmp) = \left( \begin{array}{c} \psi_{\bmp} \\ 0 \end{array} \right)\;\;\;,
\;\;\;
v(\bmp) = \left( \begin{array}{c} 0 \\ (i\sigma_2)\chi_{\bmp}^*  \end{array} \right)
\,.
\end{eqnarray}
The leading term of the amplitude corresponding to the photon ordering in 
Fig.~\ref{fig:3photon} then reads
\begin{eqnarray}
i{\cal M}_0 = -i\frac{e^3Q_f^3}{4m^2}\,
\chi_{-\bmp}^T\,(-i\sigma_2)\Big[(\bm{\sigma}\bm{\epsilon}_3)
(\bm{\sigma}\hat{\bm{k}}_3)(\bm{\sigma}\bm{\epsilon}_2)
(\bm{\sigma}\hat{\bm{k}}_1)(\bm{\sigma}\bm{\epsilon}_1)
+(\bm{\sigma}\bm{\epsilon}_3)(\bm{\sigma}\bm{\epsilon}_2)
(\bm{\sigma}\bm{\epsilon}_1)
\Big]\psi_{\bmp}\,,
\nn\\
\end{eqnarray}
where
$\hat{\bm{k}}_i=\bm{k}_i/\omega_i$ 
are the normalized three-momenta of the photons.
If we add the graphs with different ordering of the photons and write the products of
Pauli matrices in terms of the antisymmetric sigma tensors we obtain for the
$f\bar{f}\to 3\gamma$ amplitude:

\begin{eqnarray}
i{\cal M}_0 = i\frac{e^3Q_f^3}{m^2}\,\chi_{-\bmp}^T\,(-i\sigma_2) 
\!\!\!&\bigg(& \!\!\frac{m}{\omega_3}
\epsilon_1^{i_1} \hat{k}_1^{i_2} \epsilon_2^{i_3} \hat{k}_2^{i_4} \epsilon_3^{i_5} \,
\sigma^{i_1 \ldots i_5}\nn\\[3mm]
&& \!\!\!\!\!\!\!\!\!\!\!\!\!\!\!\!\!
-\frac{1}{2}\Big[\,f(1,2,3)\,{\epsilon}_1^j +  g(1,2,3)\,\hat{k}_1^j
+ \{1\leftrightarrow 2\}
+\{1\leftrightarrow 3\} \Big]\sigma^j \bigg)\psi_{\bmp}\,,
\end{eqnarray}
where
\begin{eqnarray}
f(1,2,3) &=& \Big(1+(\hat{\bm{k}}_2 \hat{\bm{k}}_3)
- (\hat{\bm{k}}_1 \hat{\bm{k}}_3) - (\hat{\bm{k}}_1\hat{\bm{k}}_2) \Big)\,
(\bm{\epsilon}_2\bm{\epsilon}_3)\nn\\
&&\;\;+ (\hat{\bm{k}}_1\bm{\epsilon}_3) \,(\hat{\bm{k}}_3 \bm{\epsilon}_2  )
+ (\hat{\bm{k}}_1 \bm{\epsilon}_2 )\,(\hat{\bm{k}}_2 \bm{\epsilon}_3)
- (\hat{\bm{k}}_3 \bm{\epsilon}_2) \,(\hat{\bm{k}}_2 \bm{\epsilon}_3 )\nn\\[3mm]
g(1,2,3) &=&  \Big((\hat{\bm{k}}_3 \bm{\epsilon}_1) + (\hat{\bm{k}}_2 \bm{\epsilon}_1)\Big)\,
(\bm{\epsilon}_2\bm{\epsilon}_3)
-(\bm{\epsilon}_1\bm{\epsilon}_3)\,(\hat{\bm{k}}_3 \bm{\epsilon}_2)
-(\bm{\epsilon}_1\bm{\epsilon}_2)\,(\hat{\bm{k}}_2 \bm{\epsilon}_3)\,.
\end{eqnarray}
Note that the $\sigma^{i_1 i_2 i_3}$ terms cancel when summing over permutations of the photons.

\section{Tensor decompositions}
\label{appendix4}

\subsection{$T^{i_1\dots i_L}\,\sigma^{i_{L+1}}$}

\begin{figure}[t!]
  \vskip 0.3cm
  \epsfxsize=15cm \centerline{\epsfbox{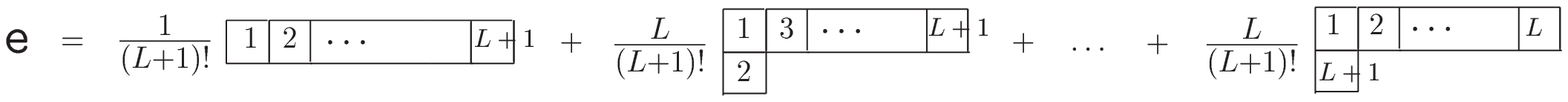}}
  \medskip
{\caption{Resolution of the identity in terms of Young 
operators of the symmetric group $S_{L+1}$ accounting also for the proper Hook
factors~\cite{Georgi}. We have only shown the Young patterns
which give a nonzero result when acting on the indices of 
$T^{i_1\dots i_L}\sigma^{i_{L+1}}$. We shall use 
$[L+1]$ to denote the first pattern, and $[L,1]$ for the 
pattern with $L$ boxes in the first row and one in the second. 
There are $L$ standard tableaux for the pattern $[L,1]$. We adopt the
convention that symmetrization horizontally 
top-to-down is followed by antisymmetrization vertically
left-to-right.} \label{fig:identity}} 
\end{figure}

The space of traceless tensors of rank $L+1$ formed from the product of the 
$T^{i_1\dots i_L}(\bmp)$ and $\sigma^{i_{L+1}}$, 
$\langle \, T^{i_1\dots i_L}\sigma^{i_{L+1}} \rangle$,
is obtained by removing the traces of $T^{i_1\dots i_L}\sigma^{i_{L+1}}$: 
\begin{eqnarray}
\langle \, T^{i_1\dots i_L}\sigma^{i_{L+1}} \rangle &=& T^{i_1\dots i_L}\sigma^{i_{L+1}} 
- \frac{1}{L+n-3} 
\sum_{k=1}^L  \delta^{i_k i_{L+1}}\Gamma^{S=1}_{L-1}(\bmp,\bmsigma)^{i_1\dots
  \ihat_k  \dots i_{L}}
\nn\\[3mm]
&& +\frac{2}{(L+n-3)(2L+n-2)}\sum_{k=1 \atop k< m}^{L+1} \delta^{i_k i_m}
\Gamma^{S=1}_{L-1}(\bmp,\bmsigma)^{i_1\dots \ihat_k \dots \ihat_m \dots i_{L+1} }
\,,
\label{Ts_traceless}
\end{eqnarray}
with the tensors $\Gamma^{S=1}_{L-1}$ defined in Eq.~(\ref{tripletcurrent2}).
This space has dimension 
$n_L\cdot n -n_{L-1}$
and is not irreducible. From the resolution of the identity in terms of Young 
operators of the symmetric group $S_{L+1}$~\cite{grouptheory} (see
Fig.~\ref{fig:identity}), 
which act on the indices of the tensors,
we can decompose 
$\langle \, T^{i_1\dots i_L}\sigma^{i_{L+1}} \rangle$ 
into traceless tensors of a given symmetry type:
\begin{eqnarray}
\langle \, T^{i_1\dots i_L}\sigma^{i_{L+1}} \rangle &=&
\frac{1}{L+1}\,\Gamma^{S=1}_{L+1}(\bmp,\bmsigma)^{i_1 \dots i_{L+1}}\nn\\[3mm]
&&+ \,\frac{L}{L+1}\,\widetilde\Gamma^{S=1}_{L}
(\bmp,\bmsigma)^{i_2 \dots i_{L}}_{[i_1 i_{L+1}]} 
-\frac{1}{L+1}\, 
\sum_{k=2}^{L}\,\widetilde\Gamma^{S=1}_{L}
(\bmp,\bmsigma)^{i_2 \dots \ihat_k \ldots i_{L+1}}_{[i_1 i_{k}]}\,.
\label{Ts_decomp}
\end{eqnarray}
The first term in Eq.~(\ref{Ts_decomp}) results from
the application of the first tableau in Fig.~\ref{fig:identity} (the pattern 
$[L+1]$) to the 
tensor $\langle \, T^{i_1\dots i_L}\sigma^{i_{L+1}} \rangle$, which yields a
totally symmetric  tensor. This is the tensor used to define the $J=L+1$
triplet current in Sec.~\ref{sectiontriplet}, and its explicit form is
given in Eq.~(\ref{tripletcurrent1}). 
The $L$ terms in the second an third lines of Eq.~(\ref{Ts_decomp}) correspond
to the $L$ Young tableaux with the pattern $[L,1]$ 
shown in Fig.~\ref{fig:identity}.
These tensors read
\begin{eqnarray}
\lefteqn{
\widetilde{\Gamma}^{S=1}_{L}(\bmp,\bmsigma)_{[i_1 i_{m}]}^{i_2\ldots \ihat_m\ldots i_{L+1}} 
  \,\equiv\,
}
\nn\\ &&
\equiv\, T^{\,i_1\dots\ihat_m\dots i_{L+1}}\,\sigma^{i_{m}}
  -  \frac{1}{L+n-3}\sum_{k=2 \atop k\ne m}^{L+1}\,
\delta^{i_k i_{m}}\,T^{\,i_1\dots\ihat_k\dots\ihat_m\dots i_{L+1} \ell}\,\sigma^{\ell}
  - \{i_1\leftrightarrow i_{m}\}\,,
 \label{newLcurrent}
 \end{eqnarray} 
and have dimension ($L\ge 1$)
\begin{eqnarray}
d_{[L,1]} \, = \,
n_L\cdot n-n_{L-1}-n_{L+1} \, = \,
\frac{\Gamma(n+L-1)}{\Gamma(n-2)\Gamma(L+2)}\frac{L(2L+n-2)}{L+n-3}\,,
\end{eqnarray} 
which for $n=3$ reduces to $2L+1$. 
Indeed, the tensors corresponding to the pattern $[L]$ in
Eq.~(\ref{tripletcurrent3}) and those corresponding to the pattern $[L,1]$ in
Eq.~(\ref{newLcurrent}), although transforming according to different  
irreducible representations of SO(n), become equivalent in $n=3$, and thus
represent both appropriate candidates 
for defining the triplet current with quantum numbers $J=L$ in dimensional
regularization. This justifies  
the quantum numbers used to label the currents in Eq.~(\ref{newLcurrent}). 

As the tensor $\langle \, T^{i_1\dots i_L}\sigma^{i_{L+1}} \rangle$ is
already symmetric on the $i_1\dots i_L$ indices, the subspaces which
result from the action of the different $[L,1]$ tableaux are all
equivalent. Therefore the space $\langle \, T^{i_1\dots
  i_L}\sigma^{i_{L+1}} \rangle$ decomposes into a subspace of totally
symmetric tensors, of dimension $n_{L+1}$, and a subspace of tensors
which transform  equivalently to the irreducible representation of
SO(n) defined by the symmetry pattern $[L,1]$, of dimension 
$d_{[L,1]}$. Thus the  second and third lines in Eq.~(\ref{Ts_decomp})
should not be understood as a tensor decomposition into orthogonal subspaces,
but as a convenient
way to write the components of a tensor of this subspace in terms of tensors of the
symmetry type given by the Young tableaux.

The results in Eqs.~(\ref{Ts_traceless}) and~(\ref{Ts_decomp})
show how the reducible tensor 
$A^{i_1\cdots i_{L+1}} = \psi_{\bmp}^\dagger \,T^{i_1\dots
  i_L}(\bmp)\sigma^{i_{L+1}}(i\sigma_2)\chi_{-\bmp}^*$  
can be fully
written in terms of the currents defined from the tensors 
$\Gamma^{S=1}_{L+1}\,,\widetilde{\Gamma}^{S=1}_{L}$ and $\Gamma^{S=1}_{L-1}$.
The latter current already emerges in Eq.~(\ref{Ts_traceless}) when
  eliminating the traces from  $T^{i_1\dots i_L}\sigma^{i_{L+1}}$.

\subsection{$T^{i_1\dots i_L}\,\sigma^{i_{L+1}i_{L+2}}$}

The reduction of the tensor  $T^{i_1\dots i_L}(\bmp)\sigma^{i_{L+1} i_{L+2}}$
can be carried out in a similar way. The traceless tensor 
$\langle \, T^{i_1\dots i_L}(\bmp)\sigma^{i_{L+1} i_{L+2}}\, \rangle$, of
dimension $n_L\frac{n(n-1)}{2}-n_{L-1}\cdot n+n_{L-2}$, 
reads
\begin{eqnarray}
\langle \, T^{i_1\dots i_L}\sigma^{i_{L+1} i_{L+2}} \rangle &=& 
T^{i_1\dots i_L}\sigma^{i_{L+1} i_{L+2}}
-c_1\sum_{k=1 \atop k<m}^L\,\delta^{i_k i_{m}}\,
\widetilde{\Gamma}^{S=1}_{L-1}(\bmp,\bmsigma)^{i_1 \dots \ihat_k \ldots \ihat_m \ldots
i_{L}}_{[i_{L+1}i_{L+2}]}
\nn\\
&&-c_0(1-\delta_{L,1})\, \bigg[ \sum_{k=1}^L\,\delta^{i_k i_{L+1}}\,
\Gamma^{S=1}_{L}(\bmp,\bmsigma)^{i_1 \dots \ihat_k \ldots i_L i_{L+2}}-\{i_{L+1}\leftrightarrow i_{L+2}\}\bigg]
\nn\\[2mm]
&&-(c_0+c_2)\,\bigg[\sum_{k,m=1 \atop k\ne m}^L\,\delta^{i_k i_{L+1}}\,
\widetilde{\Gamma}^{S=1}_{L-1}(\bmp,\bmsigma)^{i_1 \dots \ihat_k \ldots \ihat_m \ldots i_{L}}_{[i_{m}i_{L+2}]}
-\{i_{L+1}\leftrightarrow i_{L+2}\}\bigg]\,,\nn\\
\label{Tss_traceless}
\end{eqnarray}
with 
$$
c_0=\frac{1}{(L-1)(L+n-2)} \; \;,\;  
\; c_1=-\frac{2}{(2L+n-2)\,(L+n-4)}\,,
$$
$$
c_2=\frac{2}{(2L+n-2)\,(L+n-4)\,(L+n-2)}\,,
$$
and
\begin{eqnarray}
\widetilde{\Gamma}^{S=1}_{L-1}(\bmp,\bmsigma)_{[i_1 i_{m}]}^{i_2\ldots \ihat_m\ldots i_{L}} 
  \, \equiv \, T^{\,i_1\dots\ihat_m\dots i_{L}\ell}\,\sigma^{\ell\, i_{m}}
  - T^{\,i_2\dots i_{L}\ell}\,\sigma^{\ell \, i_{1}}\,.
 \label{newL-1current}
\end{eqnarray} 
The latter is a tensor which transforms according to the representation defined by the
pattern $[L,1]$, so for $n=3$ it is equivalent to the totally symmetric tensor 
$\Gamma^{S=1}_{L-1}(\bmp,\bmsigma)^{i_1\dots i_{L-1}}$ defined in
Eq.~(\ref{tripletcurrent2}) with $L-1$ indices.

The space $\langle \, T^{i_1\dots i_L}(\bmp)\sigma^{i_{L+1} i_{L+2}}\, \rangle$
can be written in terms of tensors with the symmetry type given by the standard
Young tableaux which appear in the resolution of the identity
of the symmetric group $S_{L+2}$ shown in Fig.~\ref{fig:identityss}:
\begin{figure}[t!]
  \vskip 0.3cm
  \epsfxsize=15cm \centerline{\epsfbox{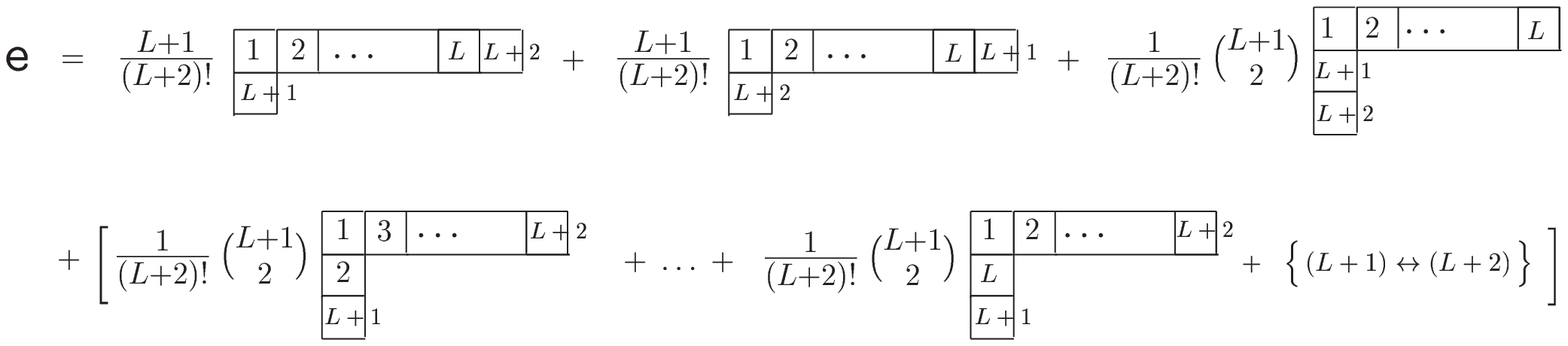}}
  \medskip
{\caption{Resolution of the identity in terms of Young 
operators of the symmetric group $S_{L+2}$. We have only shown the Young patterns
which give a nonzero result when acting on the indices of 
$T^{i_1\dots i_L}\sigma^{i_{L+1} i_{L+2}}$. There are in total $2(L-1)$ 
tableaux in the second line.} \label{fig:identityss}}
\end{figure}
\begin{eqnarray}
\langle \, T^{i_1\dots i_L}\sigma^{i_{L+1} i_{L+2}} \rangle &=&
\frac{1}{L+2}\,
\widetilde{\Gamma}^{S=1}_{L+1}
(\bmp,\bmsigma)^{i_2 \dots \ihat_{L+1} i_{L+2}}_{[i_1 i_{L+1}]}
-\frac{1}{L+2}\, 
\widetilde{\Gamma}^{S=1}_{L+1}(\bmp,\bmsigma)^{i_2 \dots i_{L+1}}_{[i_1 i_{L+2}]}+\dots\,
\label{Tss_decomp}
\end{eqnarray}
The first and second terms in Eq.~(\ref{Tss_decomp}) correspond to the
action of the two Young tableaux with 
the pattern $[L+1,1]$ in Fig.~\ref{fig:identityss}. The explicit form
of these tensors read   
\begin{eqnarray}
\widetilde{\Gamma}^{S=1}_{L+1}(\bmp,\bmsigma)^{i_2 \dots i_{L+1}}_{[i_1 i_{L+2}]} 
  &=& T^{\,i_2\dots i_{L+1}}\,\sigma^{i_1i_{L+2}}
  + \sum_{k=2}^{L+1}\,T^{\,i_1\dots \ihat_k \dots i_{L+1}}\,\sigma^{i_k i_{L+2}}
 \nn\\[3mm]
  && - \,f_1\,\sum_{k<j \atop k=2}^{L+1}\,\delta^{i_k
  i_j}\,T^{\,i_1\dots\ihat_k\dots\ihat_j \dots i_{L+1}
  \ell}\,\sigma^{\ell \,i_{L+2}}  
  \nn\\[3mm]  && 
- \,f_2\,\sum_{k\ne j \atop k,j=2}^{L+1}\,\delta^{i_1
  i_k}\,T^{\,i_2\dots\ihat_k\dots\ihat_j\dots i_{L+2} 
  \ell}\,\sigma^{\ell \,i_j}
  \nn\\[3mm]
  && - \,(f_1+f_2)\,\sum_{k=2}^{L+1}\,\delta^{i_1
  i_k}\,T^{\,i_2\dots\ihat_k\dots i_{L+1} \ell} \,\sigma^{\ell \,i_{L+2}}
 - \{i_1\leftrightarrow i_{L+2}\} \,,
  \nn\\[3mm]
  f_1 =&&\!\!\!\!\!\!\!\frac{2}{2L+n-2} \quad,\quad f_2=\frac{2L+n}{(L+n-2)(2L+n-2)}\;.
  \label{newL+1current}
\end{eqnarray}
The dots in Eq.~(\ref{Tss_decomp}) stand for the tensors associated to the
rest of the Young tableaux in Fig.~\ref{fig:identityss}. We do not give their
explicit form because they are evanescent and vanish 
for $n=3$, as the traceless tensors corresponding to Young patterns in which
the sum of the lengths  of the first two columns exceeds $n$ are identically
zero~\cite{grouptheory}~\footnote{
For the case $L=1$,   
the third diagram in Fig.~\ref{fig:identityss} does not vanish in $n=3$, but
yields a totally antisymmetric tensor.
}.   
Eqs.~(\ref{Tss_traceless}) and~(\ref{Tss_decomp}) give the decomposition 
of the tensor
$B^{i_1\cdots i_{L+2}} = \psi_{\bmp}^\dagger \,T^{i_1\dots
  i_L}(\bmp)\sigma^{i_{L+1} i_{L+2}} 
(i\sigma_2)\chi_{-\bmp}^*$  into the currents built from the irreducible
tensors $\widetilde{\Gamma}^{S=1}_{L\pm 1}$ and
$\Gamma^{S=1}_{L}$, that have fixed quantum numbers in $n=3$.

The three new spin triplet currents $\psi_{\bmp}^\dagger
\,\widetilde{\Gamma}^{S=1}_J(i\sigma_2)\chi_{-\bmp}^*$  
introduced above, Eqs.~(\ref{newLcurrent},\ref{newL-1current},\ref{newL+1current}),  
have $2J+1$ independent components in $n=3$ dimensions
and differ from the basis of currents defined in Sec.~\ref{sectiontriplet} in
the symmetry patterns and the number of indices. The following relations 
among  the tensors $\Gamma^{S=1}_J$  and 
$\widetilde{\Gamma}^{S=1}_J$  with the same quantum numbers hold for $n=3$:
\begin{eqnarray}
 \widetilde{\Gamma}^{S=1}_{L+1}(\bmp,\bmsigma)^{i_2\dots i_{L+1}}_{[j_1 j_{L+2}]}
  &=& 
i\,\frac{L+2}{L+1} \,\epsilon^{\, i_1 j_1 j_{L+2}}\,
  \Gamma^{S=1}_{L+1}(\bmp,\bmsigma)^{i_1\dots i_{L+1}} \,, 
  \nn\\[3mm]
 \widetilde{\Gamma}^{S=1}_{L}(\bmp,\bmsigma)^{i_2\dots i_{L}}_{[j_1 j_{L+1}]}
  &=& 
\frac{i}{L} \,\epsilon^{\,i_1 j_1 j_{L+1} } \,
  \Gamma^{S=1}_{L}(\bmp,\bmsigma)^{i_1\dots i_{L}}\,,  
  \nn\\[5mm] 
 \widetilde{\Gamma}^{S=1}_{L-1}(\bmp,\bmsigma)^{i_2\dots i_{L-1}}_{[j_1 j_{L}]} 
  &=& 
-i\,\epsilon^{\, i_1 j_1 j_{L}} \,\Gamma^{S=1}_{L-1}(\bmp,\bmsigma)^{i_1\dots
  i_{L-1}} \,.
\end{eqnarray}
For completeness, we also give the total contraction of the
$\widetilde{\Gamma}^{S=1}_J$ tensors: 
\begin{eqnarray} 
 \mbox{Tr}\Big[
\widetilde
\Gamma^{S=1\,\dagger}_{L+1}(\bmp,\bmsigma)^{i_2\dots\dots i_{L+1}}_{[i_1 i_{L+2}]}
 \,
\widetilde
\Gamma^{S=1}_{L+1}(\bmq,\bmsigma)^{i_2\dots\dots i_{L+1}}_{[i_1 i_{L+2}]}
 \Big] 
 & = & \frac{2(n-2)(L+2)(2L+n)(L+n-1)}{2L+n-2} 
\nn\\[2mm] && \qquad \times\,
T_{L,n}(\bmp,\bmq)
\mbox{Tr}\,[\bmone]\,,
\nn\\[2mm]
 \mbox{Tr}\Big[
\widetilde
\Gamma^{S=1\,\dagger}_{L}(\bmp,\bmsigma)^{i_2\dots i_L}_{[i_1 i_{L+1}]}
 \,
\widetilde
 \Gamma^{S=1}_{L}(\bmq,\bmsigma)^{i_2\dots i_L}_{[i_1 i_{L+1}]}
 \Big] 
 & = &  2(n-2)\,\frac{L+n-2}{L+n-3}\,
T_{L,n}(\bmp,\bmq)
\mbox{Tr}\,[\bmone]\,,
\nn\\[2mm]
 \mbox{Tr}\Big[
\widetilde
\Gamma^{S=1\,\dagger}_{L-1}(\bmp,\bmsigma)_{[i_1 i_{L}]}^{i_2\ldots i_{L-1}} 
 \,
\widetilde
 \Gamma^{S=1}_{L-1}(\bmq,\bmsigma)_{[i_1 i_{L}]}^{i_2\ldots i_{L-1}} 
 \Big] 
 & = & 2(n-2)\,
T_{L,n}(\bmp,\bmq)
 \mbox{Tr}\,[\bmone]\,.
\end{eqnarray}
To conclude this section we mention that the guidelines of
constructing nonrelativistic currents that are irreducible under SO(n),
as shown above, can be generalized in a straightforward way to
currents containing $\sigma^{i_1\dots i_m}$ with $m>2$. For larger
$m$, the number of evenescent currents will increase.
A very helpful rule to identify these currents is the rule
that traceless tensors corresponding to Young patterns in which
the sum of the lengths  of the first two columns exceeds $n$ are
identically zero~\cite{grouptheory}.

\section{Useful integrals}
\label{appendix5}

We list in this appendix the results for the relevant integrals needed
for the determination of the UV-divergences of the three-loop correlators
in  Secs.~\ref{sectionNLLsinglet} and \ref{sectionNLLtriplet}. Note that
Greek letters  $\alpha,\beta,\gamma,\ldots$ refer to real numbers while 
$k,r$ denote positive integers. 

We define the following integrals:
\begin{eqnarray}
\mbox{I}^{(k)}[\alpha,\beta,\gamma] 
&\equiv& 
\int  \frac{d^n\bmp}{(2\pi)^n} \frac{d^n\bmq}{(2\pi)^n} \,
\frac{(\bmp.\bmq)^k}
{[\bmp^2+\delta]^\alpha\,[(\bmp-\bmq)^2]^\beta\,[\bmq^2+\delta]^\gamma}
\,,
\end{eqnarray}
\begin{eqnarray}
\mbox{I}^{(k)}[\alpha,\beta,\gamma,\rho,\sigma] &\equiv& \int  
\frac{d^n\bmp}{(2\pi)^n}  \frac{d^n\bm{\ell}}{(2\pi)^n} \frac{d^n\bmq}{(2\pi)^n} \,
\frac{(\bmp.\bmq)^{k}}
{[\bmp^2+\delta]^\alpha\,[(\bmp-\bm{\ell})^2]^\beta\,
[\bm{\ell}^2+\delta]^\gamma\,[(\bm{\ell}-\bmq)^2]^\rho\,[\bmq^2+\delta]^\sigma}
\,,
\nn\\[3mm]
\end{eqnarray}
\begin{eqnarray}
\mbox{J}[\alpha,\beta,\gamma,\rho,\sigma] &\equiv& \int  
\frac{d^n\bmp}{(2\pi)^n}  \frac{d^n\bm{\ell}}{(2\pi)^n} \frac{d^n\bmq}{(2\pi)^n} \,
\frac{1}
{[\bmp^2+\delta]^\alpha\,[(\bmp-\bm{\ell})^2]^\beta\,[\bm{\ell}^2]^\gamma\,
[(\bm{\ell}-\bmq)^2]^\rho\,[\bmq^2+\delta]^\sigma}
\,.\qquad
\end{eqnarray}
%
%
%
%
%
For the spin-independent potentials the required UV-divergent terms are:
\begin{eqnarray}
\mbox{I}^{(k)}[1,\frac{1}{2},1]_{\rm div} &=& 
(-1)^{k+1}
\,{}_2F_1\Big(\frac{1}{2},-k,\frac{3}{2},2\Big)\,
\frac{\delta^{k+\frac{1}{2}}}{8\pi^3\,\epsilon}
 + {\cal O}(\epsilon^0)
\,,
\end{eqnarray}
\begin{eqnarray}
\mbox{I}^{(k)}[1,2-\frac{n}{2},1]_{\rm div} &=& 
(-1)^{k+1}
\,{}_2F_1\Big(\frac{1}{2},-k,\frac{3}{2},2\Big)\,
\frac{\delta^{k+\frac{1}{2}}}{16\pi^3\,\epsilon}
 + {\cal O}(\epsilon^0)
\,,
\end{eqnarray}
\begin{eqnarray}
\mbox{I}^{(k)}[1,1,0,1,1]_{\rm div}=(-1)^{k+1}
\,{}_2F_1\Big(\frac{1}{2},-k,\frac{3}{2},2\Big)\,
\frac{\delta^{k+\frac{1}{2}}}{128\pi^3\,\epsilon}
+ {\cal O}(\epsilon^0)
\,,
\end{eqnarray}
\begin{eqnarray}
\mbox{I}^{(2k)}[1,0,1,1,1]_{\rm div}=-\frac{1}{1+2k}
\,\frac{\delta^{2k+\frac{1}{2}}}{256 \pi^3\,\epsilon}+ {\cal O}(\epsilon^0) \,.
\nn
\end{eqnarray}
The spin-dependent contributions require:
\begin{eqnarray}
\mbox{J}[1,1,-k,1,1]_{\rm div} &=& (-1)^{k+1}\frac{1}{1+2k}
\,\frac{\delta^{k+\frac{1}{2}}}{128\pi^3\epsilon} + {\cal O}(\epsilon^0)
\,,
\end{eqnarray}
\begin{eqnarray}
\mbox{J}[1,-k,r,1,1]_{\rm div} &=& 
(-1)^{k-r}\frac{\Gamma(2k+2)}{\Gamma(2k-2r+4)\Gamma(2r)}
\,\frac{\delta^{k-r+\frac{3}{2}}}{256\pi^3\epsilon} + {\cal O}(\epsilon^0)
\,,
\end{eqnarray}
\begin{eqnarray}
\mbox{I}^{(0)}[1,1,-k,1,1]_{\rm div} &=& 
-\frac{\sqrt{\pi}\,\Gamma(k+1)}{\Gamma(k+\frac{3}{2})}
\,\frac{\delta^{k+\frac{1}{2}}}{256\pi^3\epsilon} + {\cal O}(\epsilon^0)
\,,
\end{eqnarray}
\begin{eqnarray}
\mbox{I}^{(0)}[1,-k,1,1,1]_{\rm div} &=&  
(-1)^{k+1}\frac{2^{2k-8}}{1+k}
\,\frac{\delta^{k+\frac{1}{2}}}{\pi^3\epsilon} + {\cal O}(\epsilon^0)
\,.
\end{eqnarray}
%
%

\end{document}